\documentclass[12pt]{iopart}

\expandafter\let\csname equation*\endcsname\relax
\expandafter\let\csname endequation*\endcsname\relax

\usepackage{epsfig}
\usepackage{booktabs}
\usepackage{amsmath}
\usepackage{amsfonts}
\usepackage{amssymb}
\usepackage[sort,compress]{cite}
\usepackage{nicefrac}
\usepackage{ltxtable}
\usepackage{siunitx}
\usepackage[T1]{fontenc}

\usepackage{units}
\usepackage{textcomp}
\usepackage{wasysym}
\usepackage{mathrsfs}

\DeclareMathOperator{\kei}{kei}
\DeclareMathOperator{\Ei}{Ei}

\newcommand{\deriv}[2]{\frac{\partial #1}{\partial #2}}
\newcommand{\integral}[3]{\int_{#1}^{#2}\mathrm{d}#3\,}
\newcommand{\funkintegral}[3]{\int_{#1}^{#2}{\mathcal{D}#3\,}}
\newcommand{\Summe}[2]{\sum_{#1}^{#2}}

\newcommand{\Eb}[0]{\epsilon_b}
\newcommand{\ga}{\gamma}
\newcommand{\hn}{h_0}
\newcommand{\hk}{h(\mathbf{k})}
\newcommand{\hq}{h(\mathbf{q})}
\newcommand{\hkt}[0]{h(\kft,t)}
\newcommand{\hr}{h(\mathbf{r})}
\newcommand{\hrt}{h(\mathbf{r},t)}
\newcommand{\kr}{k_0}
\newcommand{\koff}{K^{\text{off}}}
\newcommand{\kofff}{k^{\text{off}}}
\newcommand{\kon}{K^{\text{on}}}
\newcommand{\konn}{k^{\text{on}}}
\newcommand{\ka}{\kappa}
\newcommand{\kbt}[0]{k_BT}
\newcommand{\kft}[0]{\mathbf{k}}
\newcommand{\lnull}{l_0}
\newcommand{\NB}{N_b}

\newcommand{\q}{q_0}

\begin{document}
\title[Multiscale approaches to  protein-mediated interactions between membranes]{Multiscale approaches to protein-mediated interactions between membranes - Relating microscopic and macroscopic dynamics in radially growing adhesions}
\author{Timo Bihr\textsuperscript{1,2}, Udo Seifert\textsuperscript{2} and Ana-Sun\v{c}ana Smith\textsuperscript{1,3}}
\address{\textsuperscript{1}
Institut f\"ur Theoretische Physik and the Excellence Cluster: Engineering
of Advanced Materials, Universit\"at Erlangen-N\"urnberg, N\"agelsbachstrasse 49b, 91052
Erlangen, Germany
}
\address{\textsuperscript{2}II. Institut f\"ur Theoretische Physik, Universit\"at Stuttgart, Pfaffenwaldring 57, 70569 Stuttgart, Germany}
\address{\textsuperscript{3}Institute Ru\dj er Bo\v skovi\'c,  Division of Physical Chemistry, Bijeni\v{c}ka cesta 54, 10000 Zagreb, Croatia.}

\submitto{\NJP}
\date{\today}

\synctex=1

\pacs{87.10.Rt, 87.16.-b, 82.70.Uv}
\begin{abstract}
Macromolecular complexation leading to coupling of two or more cellular membranes is a crucial step in a number of biological functions of the cell.
While other mechanisms may also play a role, adhesion always involves the fluctuations of  deformable membranes, the diffusion of proteins and the molecular binding and unbinding. Because these stochastic processes couple over a multitude of time and length scales, theoretical modeling of membrane adhesion has been a major challenge. Here we present an effective Monte Carlo scheme within which the effects of the membrane are integrated into local rates for molecular recognition. The latter step in the Monte Carlo approach enables us to simulate the nucleation and growth of adhesion domains within  a system of the size of a cell for tens of seconds without loss of accuracy, as shown by comparison to $10^6$ times more expensive Langevin simulations.
To perform this validation, the Langevin approach was augmented to simulate diffusion of proteins explicitly, together with reaction kinetics and membrane dynamics.
We use the Monte Carlo scheme to gain deeper insight to the experimentally observed radial growth of micron sized adhesion domains, and connect the effective rate with which the domain is growing to the underlying microscopic events. We thus demonstrate that our technique yields detailed information about protein transport and complexation in membranes, which is a fundamental step toward understanding even more complex membrane interactions in the cellular context.   
\end{abstract}
\maketitle


\section{Introduction}
At the origin of many biological phenomena is cell adhesion promoted by the formation of macromolecular ensembles.
Despite intensive research over the last two decades \cite{Sackmann1996,Seifert1997,Groves2007,Brown2008,Weikl2009,Smith2009, Brown2011,Gao2011,Fenz2012,Maitre2012,James2012,Amack2012,Schwarz2013} and the pressing biological significance \cite{Grakoui1999,Dustin2010,Dustin2012,Brasch2012,Sackmann2014}, the growth of these structures in membranes is still poorly understood. Formation of adhesions involves a number of stochastic events occurring on different length and  timescales. The minimal system involves protein diffusion and formation of bonds, which occurs on characteristic times of $10^{-5}-10^{-2}$ s. These two processes  couple to fast membrane fluctuations ($10^{-9}-10^{-6}$ s). Several length scales are also involved - from nanometer separations necessary for molecular recognition to the micron-sized objects that are being grown. Moreover, molecular complexation induces membrane deformations which in turn promotes long-range cooperative effects. If all these elements are considered, difficulties in modeling the dynamics of macro-molecular scaffolding come as no surprise.

Early attempts to model the formation of macromolecular structures were related to interactions of protein-decorated membranes with underlying substrates containing the appropriate binding partners in the adhesion process.
Thereby, analogies with classical theories of growth (Stefan problem and kinetically limited aggregation) were explored \cite{Boulbitch2001,Freund2004, Shenoy2005, Gao2005}.
Other approaches focused on the role of the membrane fluctuations \cite{Raudino2010,Raudino2010a}.
Furthermore, a number of scaling laws were suggested after the analysis of the relationship between the various involved stochastic processes\cite{Gennes2003}.
However, only limited experimental confirmation has been obtained to support these arguments
\cite{Cuvelier2004,Puech2006}.

Later efforts concentrated on the construction of accurate simulation schemes that treat the membrane fluctuations explicitly.
First, dynamics of domain formation was studied by Monte Carlo approaches where furthermore the diffusion was treated by a random walk and complexation of proteins was explored through Metropolis rates \cite{Weikl2001,Krobath2007,Weikl2009}. Concomitantly, Langevin simulations \cite{Lin2006a,Lin2006,Brown2008,Reister-Gottfried2008,Reister2011}  were developed. In earlier attempts\cite{Lin2006a,Lin2006,Brown2008}, binding and unbinding was not considered, while latter efforts involved rates that are functions of the instantaneous membrane profile \cite{Reister-Gottfried2008,Reister2011}. The problem with all these methods is that only micron-sized systems could be studied for about a millisecond. Consequently, long time-scale dynamics associated with the formation of larger macromolecular structures, such as radially growing domains and diffusion-limited aggregation, remained out of reach.
To address these biologically relevant issues, significant efforts went toward developing coarse-grained simulation methods.
This resulted in mapping the problem onto lattice gas and Ising-like models \cite{Farago2008,Farago2010,Weil2010,Speck2010,Speck2011}, which is, however, accurate only in a limited range of parameters.

Here we build on the experience in coarse-graining the dynamics of nucleation of macromolecular complexes in membranes \cite{Bihr2012}.
We solve the problem of coupling time and length scales by constructing an effective Monte Carlo simulation scheme, for which we demonstrate applicability in a very broad range of parameters.
The stepping stone for our approach is the realization that there is a clear separation of time scales between membrane fluctuations and protein binding and diffusion.
This allows us to fully circumvent simulating the membrane, by incorporating its influence into effective rates for the (de)complexation of proteins.
We validate our scheme against explicit Langevin simulations \cite{Reister2011}, which themselves were shown to agree very well with experiments in the context of the nucleation \cite{Fenz2011} and the morphology of adhesion domains \cite{Reister-Gottfried2008}. 
In order to make this comparison easier, we first present the underlying theoretical model, its direct implementation into the augmented Langevin scheme and, then, the upscaling and the construction of the effective Monte Carlo scheme. Furthermore, we demonstrate the potential of the Monte Carlo scheme by simulating radially growing domains containing up to 10\textsuperscript{5} ligand-receptor bonds over several seconds, as observed in analogous experiments. This allows us to explore the membrane associated processes with very high precision, and to provide deeper understanding of the overall dynamics.

\section{Model}
\begin{figure}[h]
\centering
\includegraphics[width = 0.89\linewidth]{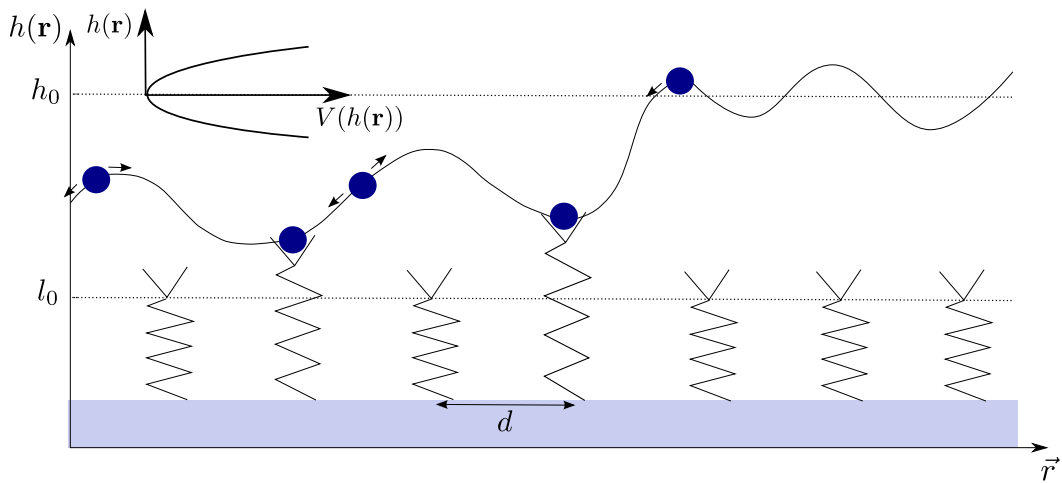}
\caption{In the model, the fluctuating membrane carries mobile ligands, which bind to immobile receptors placed equidistantly on the surface (characteristic spacing $d$). The formation of bonds is associated with the deformation of the receptor and the membrane, the latter being subject to a nonspecific harmonic potential with a minimum at $\hn$.
\label{fig:sketch}}
\end{figure}

Our model system (see figure \ref{fig:sketch}) consists of a flexible membrane that is positioned above a solid substrate. 
Receptors on the substrate can form bonds with ligands, embedded in the membrane. 
The receptors are placed on a regular square grid and are immobile in the current context. The ligands can diffuse within the membrane until a bond is formed and therefore the membrane is locally pulled towards the substrate.
More elaborate versions of our model allow for both binders to be mobile and coupled to different reservoirs, simulating a finite vesicle, or an infinite bilayer. Furthermore, binder species with different properties can be simultaneously introduced.

\subsection{The membrane}
The membrane is described as a thin sheet with an energy given by the Helfrich-Hamiltionian~\cite{Helfrich1973}
\begin{equation}
  \begin{split}
 \mathcal{H}_M [h(\mathbf{r})] &= \integral{A}{}{^2 \mathbf{r}}
  \left(\frac{\ka}{2} (\Delta h(\mathbf{r}))^2
							  + \frac{\ga}{2} \left[h(\mathbf{r})-h_0\right]^2\right)
						.
 \end{split}
 \label{eq:helfrich}
\end{equation}
Specifically, the lipid bilayer is parametrized in the Monge gauge, where the height $\hr$ is given as a  function of the position $\mathbf{r}$ of the membrane above the substrate (Fig. \ref{fig:sketch}).
A list of the variables and parameters in equation (\ref{eq:helfrich}) can be found in table \ref{tab:para}.
Specifically,  the first term in equation (\ref{eq:helfrich}) is the deformation energy of the membrane, that is itself a product of the bending rigidity $\kappa$ and the local mean  curvature of the membrane. 

While the specific protein molecules embedded in the cell wall (or
membrane) are usually considered to be responsible for cell adhesion, over the past two decades a realization emerged that the cell membrane itself, being a floppy sheet, adds another unavoidable, yet not fully understood, interaction with the opposing surface it binds to. Although this interaction does not depend  at all on any specific proteins, it can have a major impact on the protein-mediated adhesion and can be viewed as a mechanism that controls the binding affinity to the cell-adhesion molecules \cite{Bihr2014b}. Such steric interactions \cite{Helfrich1978} typically maintain the two membranes at relatively large separations $h_0$, which can be modeled by introducing a nonspecific harmonic potential of a strength~$\gamma$ with the minimum at $h_0$ \cite{Bruinsma2000,Weikl2002a,Reister2011}. The strength of this potential depends directly on the average intensity of membrane fluctuations that are themselves regulated by the tension in a membrane but also by numerous other factors such as the thickness and the composition of the glycocalyx. In the mimetic systems, this contribution is dominated by continuous interactions between the membrane and the substrate, such as gravity, Helfrich-repulsion, or Van-der-Waals forces \cite{Bruinsma2000, Schmidt2012}. The strength of this potential can be obtained experimentally by the analysis of membrane fluctuations \cite{Radler1995, Schmidt2014}.

\subsection{The bonds} 
We assume that the receptors are thermalized springs with stiffness $\lambda$ and rest length $l_0$.
This leads to the following expression for the energy of the $\NB$ bonds in the membrane
\begin{equation}
 \mathcal{H}_B [h(\mathbf{r})] = \Summe{\mathrm{i}=1}{\NB} \delta(\mathbf{r}-\mathbf{r_i}) \left[ \frac{\lambda}{2}(h(\mathbf{r})-l_0)^2-\Eb\right].
  \label{eq:helfrich2}
\end{equation}
Here, $\Eb$ accounts for the bond enthalpy gain for forming a bond and $\delta(\mathbf{r}-\mathbf{r_i})$ is the Dirac-Delta function for the positions $\mathbf{r_i}$ of the bonds.

\begin{table}
\caption{Variables and parameters of our Helfrich-Hamiltionian (see equations (\ref{eq:helfrich}) and (\ref{eq:helfrich2}))\label{tab:para}.}
\vskip 0.1cm
\centering
\begin{tabular}{clc}
\hline 
Quantity			&	Meaning								&	Unit\\
\hline 
$a$					&	lattice constant						&	$10\,\text{nm}$\\
$\kbt$				&	thermal energy at $300\,\text{K}$		&	$4.14\times 10^{-21}\,\text{J}$\\
$\ka$				& 	bending rigidity						&	$\kbt$\\
$h(\mathbf{r})$	&	membrane profile						&	$a$\\
$\ga$				&	curvature of the interaction potential	&	${\kbt}/{a^4}$\\
$\hn$				&	minimum of the interaction potential	&	$a$\\
$\NB(t)$				&	number of bonds							&	-\\
$\mathbf{r_i}$		&	position of bond $i$					&	$a$\\
$\lambda$			&	stiffness of the bond/receptor			&	${\kbt}/{a^2}$\\
$\lnull$			&	rest length of the bond/receptor		&	$a$\\
$\Eb$				&	binding enthalpy						&	$\kbt$\\
\hline 
\end{tabular}
\end{table}
 
As we assume that the structural fluctuations of free receptors occur on a faster time scale than the membrane dynamics, each bond fulfills a local detailed balance for the transitions between the bound and unbound states, given by the rates $\kofff\left(\hrt\right)$ and $\konn\left(\hrt\right)$ as
\begin{equation}
 \frac{\kofff\left(\hrt\right)}{\konn\left(\hrt\right)}=\exp{\left[ \left( \frac{\lambda}{2}\left(\hrt-\lnull\right)^2- \Eb\right)- \frac{1}{2}\ln\left(\frac{\lambda\alpha^2}{2\pi}\right)\right]}.
 \label{eq:DB}
\end{equation}
Following this condition, each bond is locally in thermal equilibrium with the instantaneous  membrane profile. Here, $\alpha$ is the range of the interaction potential of the ligand-receptor bond and for simplicity, we set $\beta=(\kbt)^{-1}\equiv 1$.
Equation (\ref{eq:DB}) depends on the stretching energy of the bond (first term in the exponent), the binding affinity (second term) and an entropic contribution (last term) which describes the  suppression of fluctuations if a receptor is bound to a ligand.
This  entropic contribution  lowers the effective binding affinity \cite{Schmidt2012}.

Inspired by \cite{Dembo1988,Erdmann2006,Gao2011} we choose the rates for the creation of a bond as 
\begin{equation}
k^{\text{on}}\left(\hrt\right)=
\kr\sqrt{\frac{\lambda\alpha^2}{2\pi}} \exp{\left[ -\frac{\lambda}{2}\left\{ \left( \hrt-\lnull   \right) -\alpha\right\}^2\right]},
\label{eq:kon}
\end{equation}
where $\kr$ is the intrinsic reaction rate.
From this local on-rate and the detailed balance condition the local, off-rate can be determined readily
\begin{equation}
 k^{\text{off}}\left(\hrt\right)=k_0 \exp{\left[- \Eb\right]}
 \exp{\left[ \lambda  \left(\hrt-\lnull\right) \alpha -\frac{\lambda\alpha^2}{2}\right]},
 \label{eq:koff}
\end{equation}
which is proportional to the rate of the Bell-model \cite{Bell1978}, accounting for the force acting on a bond $\lambda  \left(\hrt-\lnull\right)$. 

\begin{figure}
\centering
\includegraphics[scale=1]{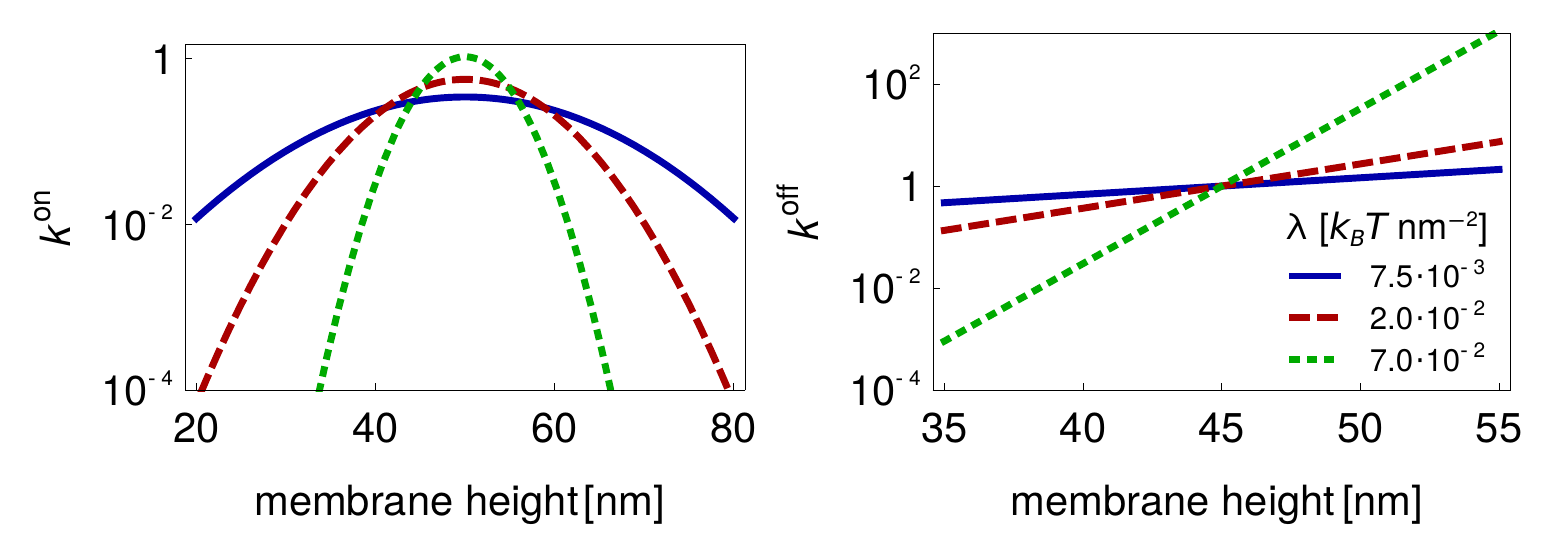}
\vskip -0.5cm
\caption{Local reaction binding (left) and unbinding (right) rates equations (\ref{eq:kon}) and (\ref{eq:koff}) shown as function of the membrane height. Other parameters were set as in table~\ref{tab:para2}, except for  the binding energy  ($\Eb=0$).
\label{fig:rates}
}
\end{figure}

We show the reaction rates eqs. (\ref{eq:kon}) and (\ref{eq:koff}) in Fig.~\ref{fig:rates}. The association rate adopts the form of a Gaussian with a width inversely proportional to the stiffness of the receptor. The maximal association rate is obtained at $l_0+\alpha$, which is at the outer edge of the potential well associated with an unperturbed receptor.
The off-rate increases exponentially with the distance between the receptor and the ligand.
Interestingly, if the ligand is in the middle of the binding region ($l_0+\alpha/2$), the stiffness of the receptor does not affect the breaking of the bond (the bond is not stressed).
\subsection{Diffusion}
Due to the membrane fluidity, the molecules within the bilayer diffuse on its surface\cite{Saffman1975}. Even though there may be an influence of the membrane elasticity on the diffusion of embedded proteins (for example by the curvature that a protein induces in the membrane \cite{Reister-Gottfried2008,Naji2009,Reister-Gottfried2010,Quemeneur2014}), these effects seem to be small for experimental relevant parameters \cite{Reister-Gottfried2010}. Therefore, we simulate the mobility of binders by a random walk, whereby  two proteins interact laterally by a hard-core potential.
The time step of the random walk is given by
\begin{equation}
\tau_d=\frac{a^2}{4D},
\label{eq:timestep_diff}
\end{equation}
with the diffusion constant $D$.
In the current work, only the ligands embedded in the membrane of the vesicle are allowed to diffuse. 
However, it is straightforward to extend the simulation scheme to situations in which both binders retain lateral mobility and explore the surface of the membrane.
The latter may be finite as in the case of vesicles and cells. 
These situations are simulated using periodic boundary conditions on the level of the system, with a selected area in the center of the simulation box representing the area of contact between two cells or the cell/vesicle and the substrate. Consequently, the formation of bonds can take place only within this region, and the remainder of the system will be depleted from the binders due to the accumulation in the zone of contact.

Binders can be also embedded in bilayers, which provides a constant chemical potential. For simulations of interactions with vesicles, a contact zone is defined, and the periodic boundary conditions are imposed for the bilayer grid.
However, to maintain the constant chemical potential (constant concentration of binders in the bulk), entering and exiting of a binder from the contact zone is associated with placing or removing a binder from a random position outside the contact zone. 

\section{Langevin simulation scheme}
\subsection {Equation of motion for the membrane}
\begin{figure}
\centering
\includegraphics[width=0.9\linewidth]{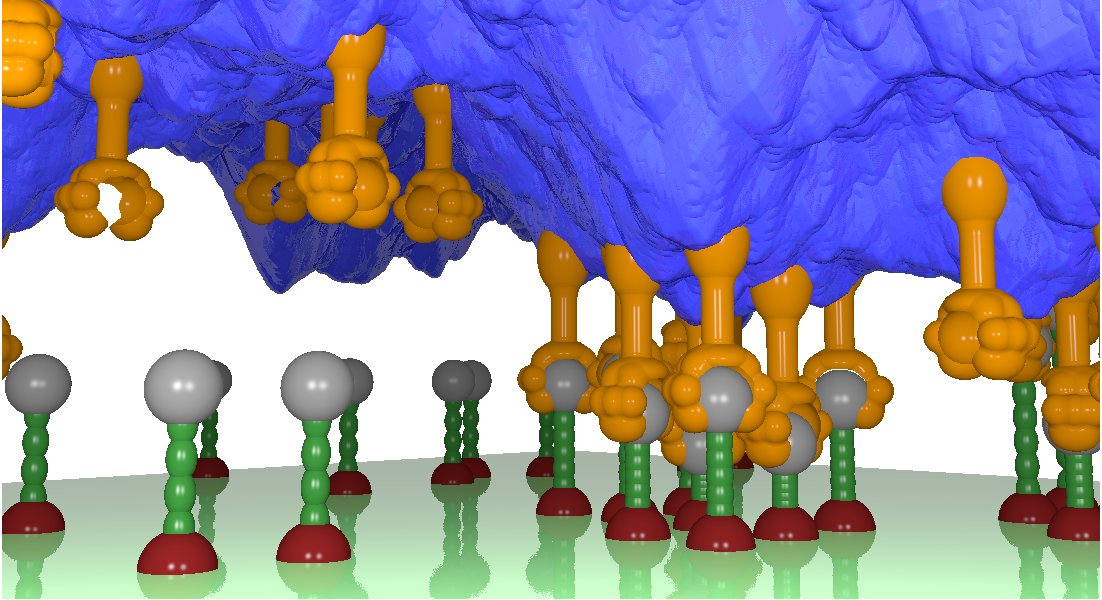}
\caption{
Snapshot of a simulation run.
The grey receptor can form bonds with the orange ligands embedded in the fluctuating membrane (blue).
The ligands can diffuse freely within in the membrane, whereas the receptors are immobilized and placed on a square grid.
\label{fig:snapshot}
}
\end{figure}

In this scheme, the membrane shape (see Fig.~\ref{fig:snapshot}) is determined explicitly in every time step.
Thereby, the system is propagated in time by means of the Langevin equation in Fourier space (see e.g. \cite{Brown2008,Reister-Gottfried2008}) derived from 
 the equations (\ref{eq:helfrich}) and (\ref{eq:helfrich2}) 
\begin{equation}
\begin{split}
\deriv{\hkt}{t}=&-\Lambda(\mathbf{k})\biggr\{ \left[\ka k^4+\ga\right](\hkt-\delta_{\mathbf{k},0} A \hn)  \\&+
\Summe{i=1}{\NB(t)} \lambda(h(\mathbf{r}_i,t)-l_0)\exp{\left(-\text{i}\mathbf{k} \cdot \mathbf{r}_i      \right)}
 \biggr\} +\xi(\mathbf{k}).
\end{split}
\label{eq:langevin}
\end{equation}
Here, $\Lambda(\mathbf{k})$ is the Oseen tensor, describing the hydrodynamic interaction between membrane and surrounding fluid and $A$ is the area of the membrane. The stochastic force $\xi(\mathbf{k})$ in the Langevin equation above is set by the temperature of the surrounding fluid. Thereby, the Oseen tensor is connected to the stochastic force by the fluctuation-dissipation theorem
\begin{equation}
\left\langle  \xi(\mathbf{k}) \xi(\mathbf{k'})  \right\rangle =  2 \kbt \Lambda (\mathbf{k}) \delta(\mathbf{k}+\mathbf{k'}).
\end{equation}
The definition of the Fourier transformation of the Langevin equation is given  by 
\begin{equation}
\hk= \integral{A}{}{^2\mathbf{r}} \exp{\left(-\text{i}\mathbf{k} \cdot \mathbf{r} \right)}h(\mathbf{r});\quad \quad \hr=\frac{1}{A} \Summe{\mathbf{k}}{}\exp{\left(\text{i}\mathbf{k} \cdot \mathbf{r}      \right)}\hk.
\end{equation}
In general, it was shown that the Oseen tensor depends on the geometry of the membrane \cite{Seifert1994}.
However, this dependence is very weak  for membranes far away from the substrate and only relevant for the four largest modes of the membrane for the parameters used in the simulations.
Thus, we use the Oseen tensor for a free membrane
\begin{equation}
\Lambda(\mathbf{k})=\frac{1}{4\eta k},
\end{equation}
where $\eta$ is the viscosity of the surrounding fluid.
The Oseen tensor for the $\mathbf{k}=0$ mode  diverges.
Following \cite{Brown2008}, the Oseen tensor  for this mode is set to
\begin{equation}
\Lambda(\mathbf{k})=\frac{3\sqrt{A}}{8\pi \eta}.
\end{equation}
The Langevin-equation (\ref{eq:langevin}) is solved numerically with the help of the Euler-Maruyama scheme (see for example \cite{Gardiner1985}).
The time step in this scheme has to be set below the smallest time scale of the membrane
\begin{equation}
\tau(k_{\text{max}})  = \frac{4\eta k_{\text{max}}}{\ka k_{\text{max}}^4 + \ga}
,
\end{equation}
which is the  typical relaxation time of the  mode with the largest $\mathbf{k}$  in the simulation ($k_{\text{max}}=\sqrt{2}\pi$)
and is on the order of $10^{-9}$ s.

\subsection{Simulation scheme}
\begin{figure}
\centering
\includegraphics[width=0.95\linewidth]{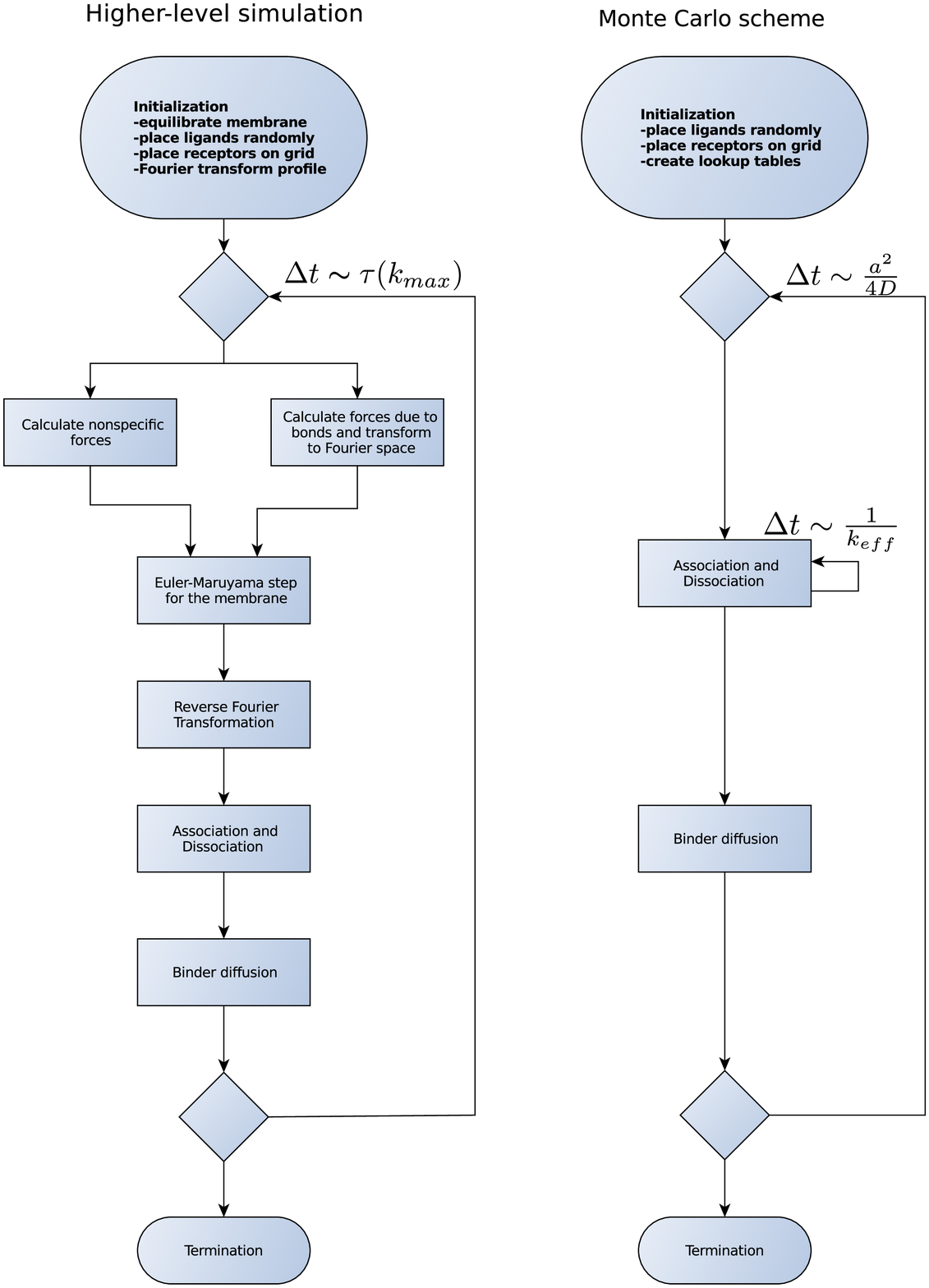}
\caption{Simulation schemes.
	Left side: Langevin scheme. The membrane is simulated explicitly. However, computational expensive Fourier transformations have to be performed.
	Right side: Effective Monte Carlo scheme. The binding kinetics is simulated with the effective reaction rates (\ref{eq:keff}) and circumventing the explicit treatment of the membrane.
	\label{fig:fcfull}	}
\end{figure}

The simulation is performed following the algorithm shown on the left in figure \ref{fig:fcfull}.
The first step initializes the system. This involves the thermal equilibration of a free membrane obtained by executing  $10^6$ steps in the time loop explained below without the reactions and binder diffusion. After that, the ligands are placed randomly on their lattice, and the receptors are put on a grid of the second lattice.

The second step is the initialization of the time loop, where the step accounts for the shortest characteristic membrane time scale ($\Delta t\equiv \tau(k_{\text{max}})$). Every time step involves (i) the calculation of the force on the membrane induced by the formed bonds in real space; (ii) the transformation of this force to the Fourier space; and (iii) the determination of bending and unspecific forces in Fourier space (first term in equation~(\ref{eq:langevin})). The sum of this forces is input to the Euler-Maruyama step, within which the membrane profile is updated in Fourier space and
transformed back to real space. 
This back-transformation is a prerequisite for the execution of the association and dissociation step. 
Here, the binding probabilities are obtained from the equations (\ref{eq:kon}) and (\ref{eq:koff}), in which the height of the membrane  and the time step of the simulation are required.
As the time scale of the reactions is much larger than the typical time scale of the membrane, these probabilities are rather small.

Finally, the mobile ligands need to be displaced to one of the neighboring  unoccupied sites. In principle, the diffusion of binders is characterized by the time step given by equation (\ref{eq:timestep_diff}), in which case the probability to jump in any direction would be  $1/4$.
However, as the time scale of the diffusion is typically several orders of magnitude larger than the step of the time loop (i.~e. $\tau_D \gg \tau(k_{\text{max}})$), the probability of a jump is rescaled to
\begin{equation}
p=\frac{\tau(k_{\text{max}})}{4\tau_D}.
\end{equation}
This new probability guarantees the correct diffusive behaviour of the ligands.

In the current simulations, the ligands are immobilized after they form a bond with a receptor, which means that only free ligands diffuse. This restriction is motivated by the experimental observation that the bonds change position only if they are subject to a significant lateral force \cite{Smith2008}.
After the diffusion has been resolved, a new iteration in the time loop is started, or the simulation is terminated.

Computationally, most time in this simulation scheme is consumed by Fast Fourier Transformations of the membrane profile and the forces, which scale like $N\log(N)$, ($N$ is the number of considered lattice points), and not linearly like other operations (diffusion and reaction kinetics).
Furthermore, the time step has to be chosen very small to accurately describe the time evolution of the membrane, and a large number of replicas must be produced to obtain a statistically sound representation of the system. These are the main reasons which make this simulation scheme computationally very expensive allowing only for length scales of up to $1\,\mu\text{m}^2$ to be simulated for up to $10^{-1}\, \text{s}$.
\begin{table}
\caption{Parameters used  in the simulations  \label{tab:para2}}
\centering
\begin{tabular}{lll}
\hline 
Parameter			&	value						\\
\hline 
$a$					&	$10\, \text{nm}$							\\
$\ka$				& 	$10\,	\kbt$								\\
$\ga$				&	$3.125\times 10^{-3}\,{\kbt}/{a^4}$			\\
$\lambda$			&	$7.5\times 10^{-3}\,\,\,\text{to}\,\,\,5\times 10^{-2}\, {\kbt}/{\text{nm}^{2}}$						\\
$\eta$				&	$2.4\times 10^{-7}\, \kbt \text{s} a^{-3}$	\\
$\hn$				&	$8\, a$										\\
$\lnull$			&	$4\,a$										\\
$d$					&	$8\,a$										\\
$\rho_l$			&	$1/64 a^{-1}$								\\
$\kr$				&   $2\times10^4\,\,\,\text{to}\,\,\,5\times10^5 \text{s}^{-1}$						\\
$\Eb$				&	$5\,\,\, \text{to} \,\,\,10\,\kbt$						\\
$D$					&	$5\times 10^5\,\,\, \text{to}\,\,\,5\times 10^7\, \text{nm}^2/ \text{s}$						\\
\hline 
\end{tabular}
\end{table}
\section{Effective Monte Carlo simulation }
\label{sec:effMonte}
\subsection{Effective rates}
The difficulties that arise with Langevin simulations could be circumvented if the explicit treatment of the membrane could be avoided. We achieve this goal in an effective Monte Carlo scheme which is based on the recently acquired understanding of the effects of the  membrane on the formation of bonds \cite{Bihr2012,Schmidt2012,Schmidt2014}. 
This scheme relies on the fact that the typical time scale of the membrane fluctuations depends on the viscosity $\eta$ of the surrounding fluid 
\begin{equation}
\tau_{\text{mem}}=\frac{4\eta \q}{\kappa \q^4+\ga}=\frac{2 \eta \q}{\ga}\simeq 2 \times 10^{-5}\,\text{s}.
\end{equation}
Here, $\q=(\gamma/\kappa)^{1/4}$ is the inverse lateral correlation length for a membrane without tension. 

Importantly, even the slowest modes are significantly faster than the reaction kinetics for ligand-receptor binding (the fastest avidin-biotin in membranes was reported to take place at $\sim 10^{3}\,\text{s}^{-1}$ \cite{Fenz2011,Bihr2014b}), while other pairs are found at $\sim 10^{2}\,\text{s}^{-1}$ \cite{Gao2011,Bihr2014b}.
Consequently, the membrane fluctuations can be regarded as equilibrated with fixed mean shape as long as the configuration of bonds interacting with the membrane remains unchanged. During this time, the fluctuating membrane, and with it the ligands, sample the entire probability distribution of distances between ligands and receptors.
In the following, we denote the height distribution at the considered binding site  $\mathbf{r}$ before the bond has formed by $p(h^r)$, and the height distribution after a bound ligand-receptor pair is formed  by  $p(h^b)$. The latter is non-trivial if the receptor or the bond itself maintains some flexibility.

The first and the second moment of these typically Gaussian distributions can be calculated explicitly for an arbitrary bond configuration \cite{Schmidt2012}. Specifically, we calculate a functional integral over all membrane profiles weighted by their Boltzmann factor (see \ref{sec:mhd} for technical details). As  result, we obtain the mean height
\begin{equation}
 \label{eq:expect2}
 \langle h(\mathbf{r})\rangle\equiv
 {\bar{h}^{r/b}}(\mathbf{r})=
  \frac
			      {-\frac{4}{\pi}  \Summe{ij}{} \lnull G_{ij}(\mathbf{r})^{-1} \kei{\left( \q \ |\mathbf{r} - \mathbf{r_i}|\right)}}
			      {8 \sqrt{\ka \ga}  + \frac{16}{\pi^2}  \Summe{ij}{} \kei{\left( \q \ |\mathbf{r} - \mathbf{r_i}|\right)}  G_{ij}(\mathbf{r})^{-1} \kei{\left( \q \ |\mathbf{r} - \mathbf{r_j}|\right)}    }
\end{equation}
and the fluctuation amplitude
\begin{equation}
 \label{eq:fluc2}
 \begin{split}
   &\langle h^2(\mathbf{r}) \rangle - \langle h(\mathbf{r}) \rangle^2  \equiv 
   \left(\sigma^{r/b}\right)  ^2(\mathbf{r})=\\&
   \left(8 \sqrt{\ka \ga} +\Summe{ij}{}
			\frac{16 \kei\left( \q \ |\mathbf{r} - \mathbf{r_i}|\right)G_{ij}(\mathbf{r})^{-1}\kei\left( \q \ |\mathbf{r} - \mathbf{r_j}|\right)}{\pi^2}  \right)^{-1}.
 \end{split}
\end{equation}
The sum runs over all pairs of bonds in the membrane at the positions $\mathbf{r_i}$ and $\mathbf{r_j}$, while $\kei(x)$ is the Kelvin function \cite{Gradshteyn2007}. The elements of the coupling matrix $G_{ij}(\mathbf{r})$ are the effects of the existing bonds on the shape and fluctuations at the arbitrary position $\mathbf{r}$, whereby the membrane mediated interaction between the bonds are comprised in the off-diagonal elements (see \ref{eq:GIJ} for the explicit form of the matrix).

\begin{figure}
\centering 
\includegraphics[scale=1]{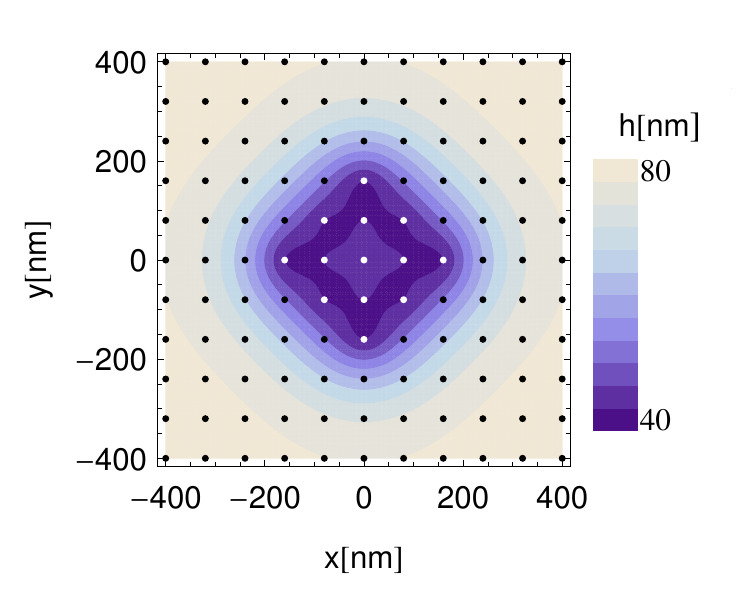}
\includegraphics[scale=1]{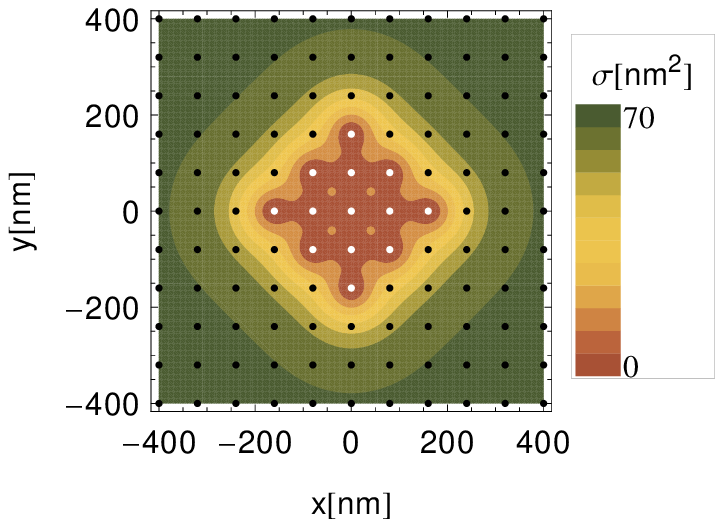}
\includegraphics[scale=1]{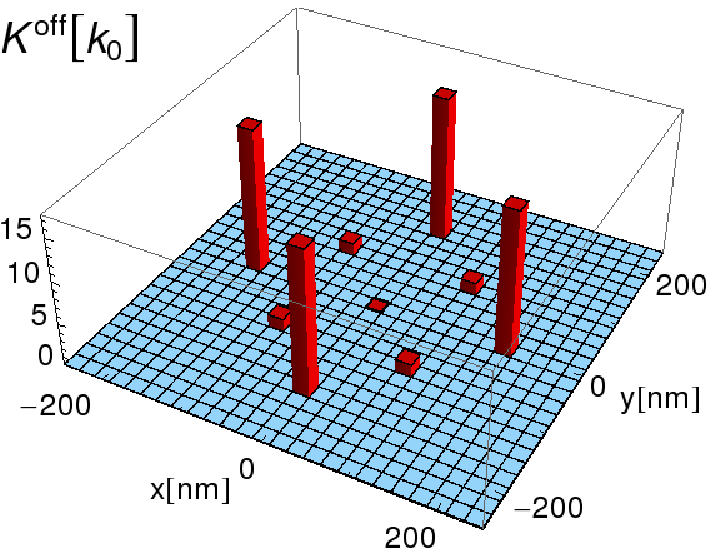}
\includegraphics[scale=1]{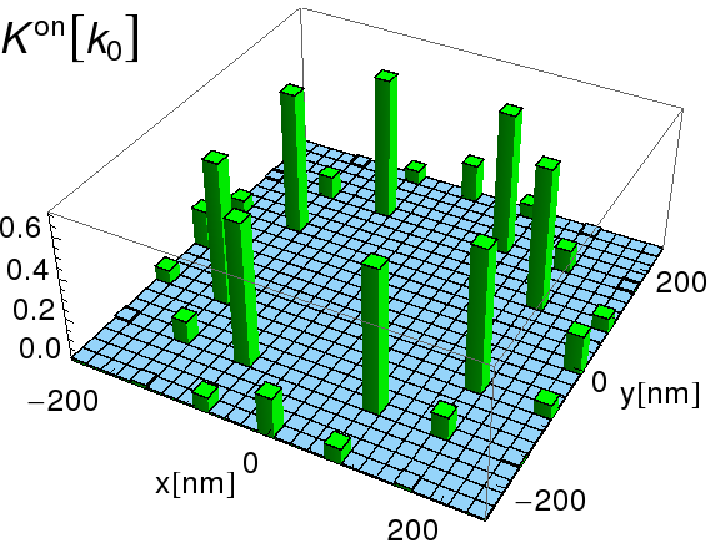}
\caption{
Membrane profile (top, left), fluctuations (top, right), effective off-rates (bottom, left) and effective on-rates (bottom, right) for a given bond configuration (white bonds and black free binding sites). The bond stiffness $\lambda$ is set to infinity  and the binding affinity $\Eb$ to zero for simplicity. The remaining parameters can be found in table \ref{tab:para2}.
\label{fig:height_fluc_map}
}
\end{figure}

The average shape and the fluctuation amplitude of a membrane containing a small cluster of bonds are shown in the top panels of figure \ref{fig:height_fluc_map}. At large distances from the cluster, the  membrane is on average flat since it resides and fluctuates in the minimum of the nonspecific potential. Because of a relatively high concentration of bonds within the cluster the membrane is likewise flat on average, but much closer to the  substrate. At the same time, its fluctuations are strongly suppressed. However, the shape and fluctuations of the membrane are significantly different in the vicinity of the bonds at the edge and in the center of the cluster. 

We use the two height distribution functions to average the Bell-Dembo rates (eq. (\ref{eq:kon}) and (\ref{eq:koff})) at the position of a free or a bound receptor giving rise to effective binding and unbinding rates   
\begin{equation}
\begin{split}
  \kon & \equiv
    \integral{}{}{h^r}p(h^r) \konn(h^r)\,\,\,\\
   \koff & \equiv \integral{}{}{h^b}p(h^b) \kofff(h^b).
\end{split}
\label{eq:intkon}
\end{equation}
Appropriately inserting equations (\ref{eq:kon}), (\ref{eq:koff}),  (\ref{eq:expect2}), and (\ref{eq:fluc2}) into the the above expression, and evaluating the integrals yields 
\begin{equation}
\begin{split}
 \kon  &= \kr \frac{\sqrt{\lambda\alpha^2} }{\sqrt{2 \pi (1+\lambda\left(\sigma ^{r} \right)^2)}}\exp{\left[
      \frac{\lambda\left[ \bar{h}^{r}-(\alpha+\lnull)   \right]^2}{2(1+\lambda\left(\sigma ^{r} \right)^2)}\right] },\\
 \koff  &= \kr \exp{\left[ \left(\frac{\lambda \alpha}{2}  \{ 2 \left(  \bar{h}^{b}-\lnull\right)+\alpha [\lambda \left(\sigma ^{b} \right)^2-1]\}-\Eb\right)
 \right] }.
 \label{eq:keff}
\end{split}
\end{equation}
In this manner, the effective rates describing the association and the dissociation of a bond depends on the exact position of a bond, and the time dependent configuration of all bonds in the system. Examples of such rates for one bond configuration can be seen in bottom panels of figure \ref{fig:height_fluc_map}. Obviously, the rates reflect the average shape and fluctuations within the membrane (top panels), which are the result of the bond configuration around the respective binding site. 
The dissociation rate of a bond at the rim is up to two orders of magnitude larger than for a bond deep within the domain (see figure \ref{fig:height_fluc_map}). 
This is due to the stabilization effects  of the neighboring bonds, which share the deformation load and cooperatively suppress the fluctuations. On the other hand, the association is the largest near the bond domain and exponentially decreases on the length scale of the lateral correlation length with increasing distance to the domain.

\subsection{Simulation scheme}

\begin{figure}
\centering
\includegraphics[width=0.95\linewidth]{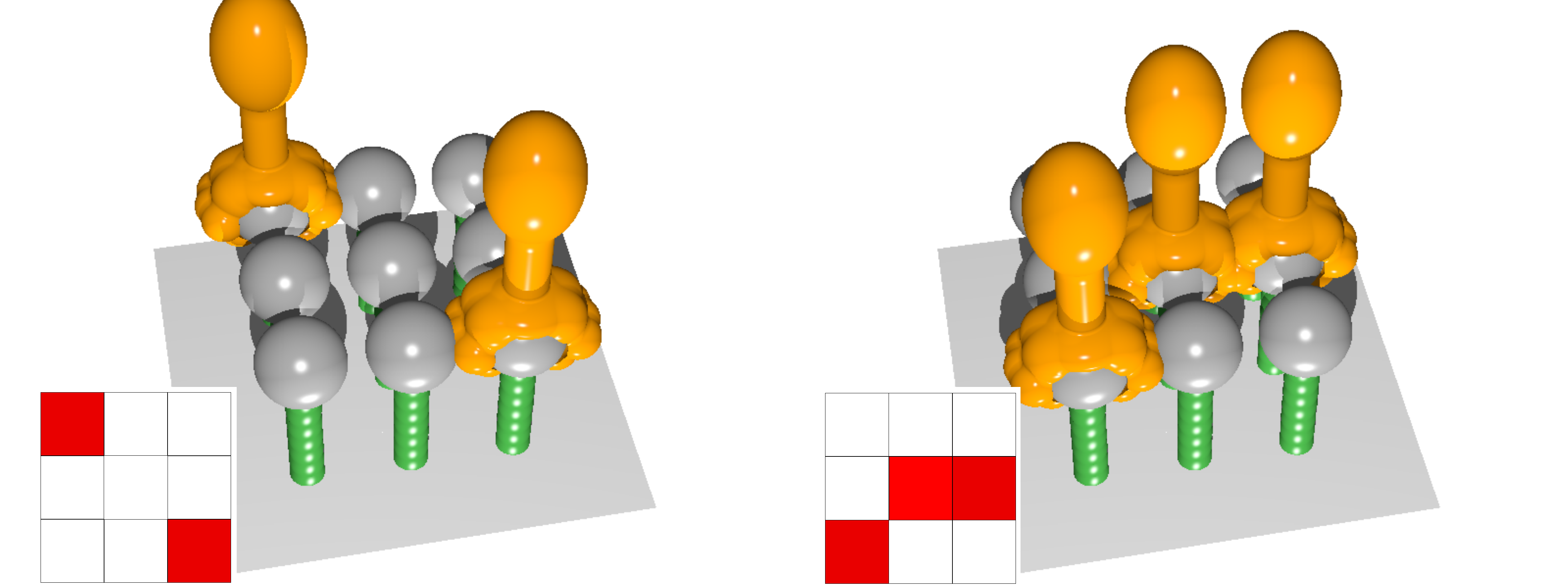}
\caption{
Example for a possible bond configuration two (left) or three (right) bonds (red squares). The association (left) or dissociation (right) rate  at the considered binding site (center square) is determined by identifying the bond configurations (red) around the binding site and retrieving the appropriate reaction rate from the lookup table. This is done for all binding sites during one iteration.
\label{fig:lookup}}
\end{figure}

We can now construct a Monte Carlo simulation of the adhesion process in which only the reaction kinetics and the diffusion of binders must be treated explicitly. Thereby, the effective rates for breaking or forming a bond at the given site must be determined for each site in every time step. 
This in turns requires inverting the coupling matrix containing all bonds, for every site in every step. In order to make the simulation fast, we assume that only the first two sets of neighboring bonds affect the rates on a particular (un)binding site (figure \ref{fig:lookup}). Consequently, only the configuration of bonds in the immediate environment is taken into account in the calculation of the effective rates. Since this environment consists only of 9 sites, all possible configurations can be explored \textit{a-priori}, and their respective rates used to create a lookup table. This restriction to the next-nearest neighbours is justified because the binding rates decay very fast with increasing distance between the bonds.

A flow chart of the Monte Carlo scheme is shown on the right panel of figure~\ref{fig:fcfull}. To initialize the system, all ligands and receptors are positioned on their respective grids as in the  Langevin simulations (random and ordered distributions are possible). Furthermore, the characteristic time steps are determined. The time step of the simulation is given by the characteristic diffusion time $\Delta t_D$ (equation (\ref{eq:timestep_diff})).
 The time step for the reaction kinetics is set to be $\Delta t_B=\Delta t_D/n$, where $n$ is the smallest integer satisfying the inequality $K^{\text{on/off}}\Delta t_D/n<1$.
  From here the probabilities for binding and unbinding are calculated as $K^{\text{on/off}} \Delta t_B$, and stored in a lookup table.
 
The simulation step starts with the reaction loop which consists of $n$ iterations. In each iteration, for every binding site (i) the bond configuration is determined, (ii) the appropriate rate is retrieved from the lookup table, (iii) association or dissociation is attempted, and (iv) the bond configuration is updated. Following the reaction loop, each binder attempts to move to a neighboring site in a same manner as in the Langevin scheme. This completes the simulation step and the system is propagated in time until the program is terminated.
While  the program allows for the diffusion of both binder types, the following discussion will be restricted to the case when the receptors are immobilized.

The advantage of the Monte Carlo scheme is that it allows for a larger time step and avoids Fast Fourier Transformations limiting the Langevin code. This allows us  to simulate length scales of several tens of micrometers and time scales of several seconds with the resolution of about 100 nm  and 10\textsuperscript{-5} s, which is necessary to understand biological processes.

\section{Validation of the Monte Carlo scheme}
\label{sec:comp}

In order to evaluate the applicability of the effective rates, we perform an extensive comparison of the results of the Langevin and Monte Carlo simulations. For this purpose, all parameters, the system size, and the statistics of data acquisition  in the two approaches is identical and no fit parameters are used in the following discussion.  

We explore a very wide range of parameters: from soft to rather stiff receptors, binding affinities from the unstable to the enthalpy dominated adhesion, fast and slow diffusion of ligands (equivalent to changing the attempt reaction frequency).

\subsection{Early stages of domain formation - nucleation dynamics}

\begin{figure}[t]
\centering
\includegraphics[scale=1]{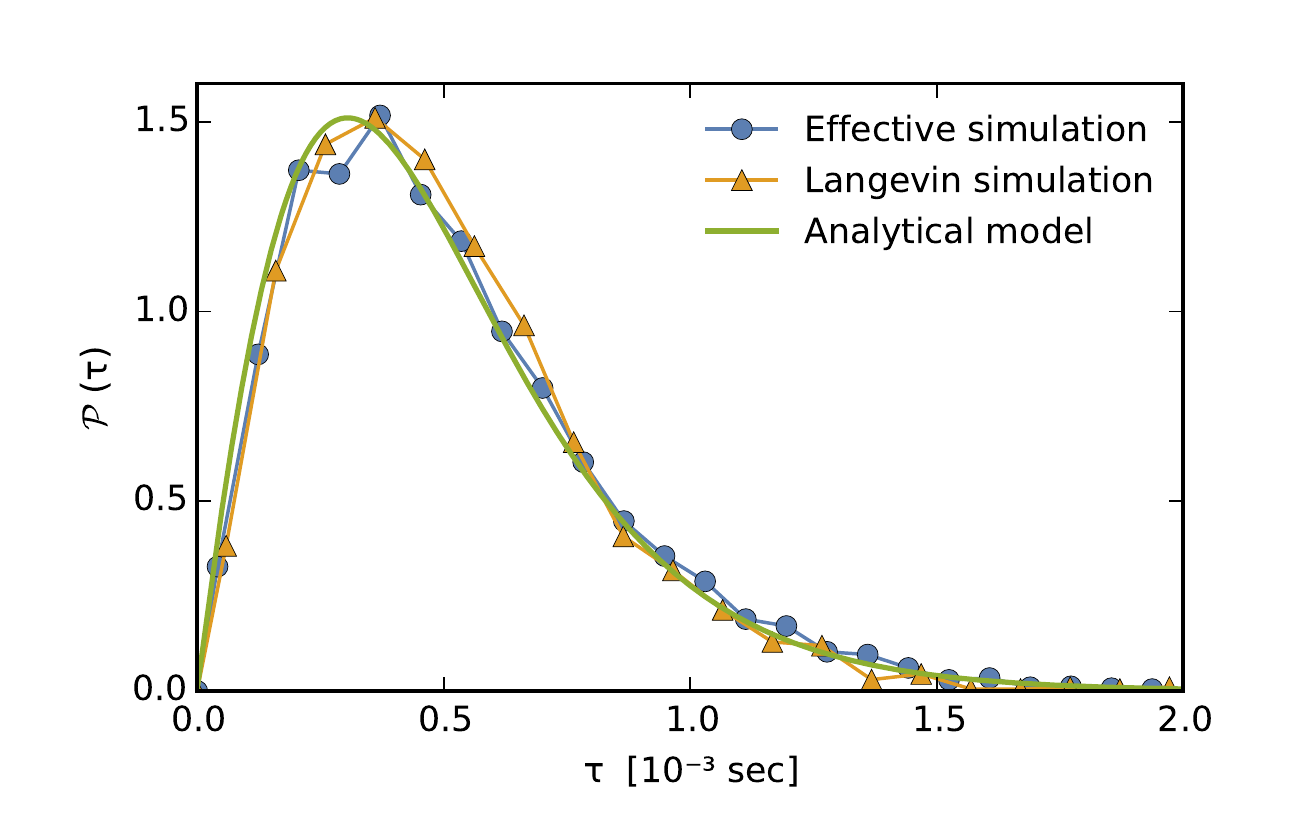}
\caption{Distribution of nucleation times.
The effective scheme, the Langevin scheme and the analytical model produce the same distribution of nucleation times (without fitting parameter).
The analytical curve is determined from the equation (6) in Bihr et al. \cite{Bihr2012}.
The intrinsic binding affinity for protein binding is set to $\epsilon_b=6.56\,\kbt$, while the diffusion constant is $D=5\,\mu \text{m}^2/\text{s}$.
All simulations were performed in a simulation box of $640\, \text{nm}\times 640\,\text{nm}$ with the densities of receptors and ligands of $\rho_r=\rho_b=1.5625\times 10^{-4} \,\text{nm}^{-2}$. The intrinsic binding rate was set to $k_0 = 10^5\,\text{s}^{-1}$,  and  the receptors are modeled as springs of stiffness $\lambda=2\times 10^{-2}\,k_BT/\text{nm}^2$.
\label{fig:nucleation}}
\end{figure}

\begin{figure}
\centering
\includegraphics[scale=1]{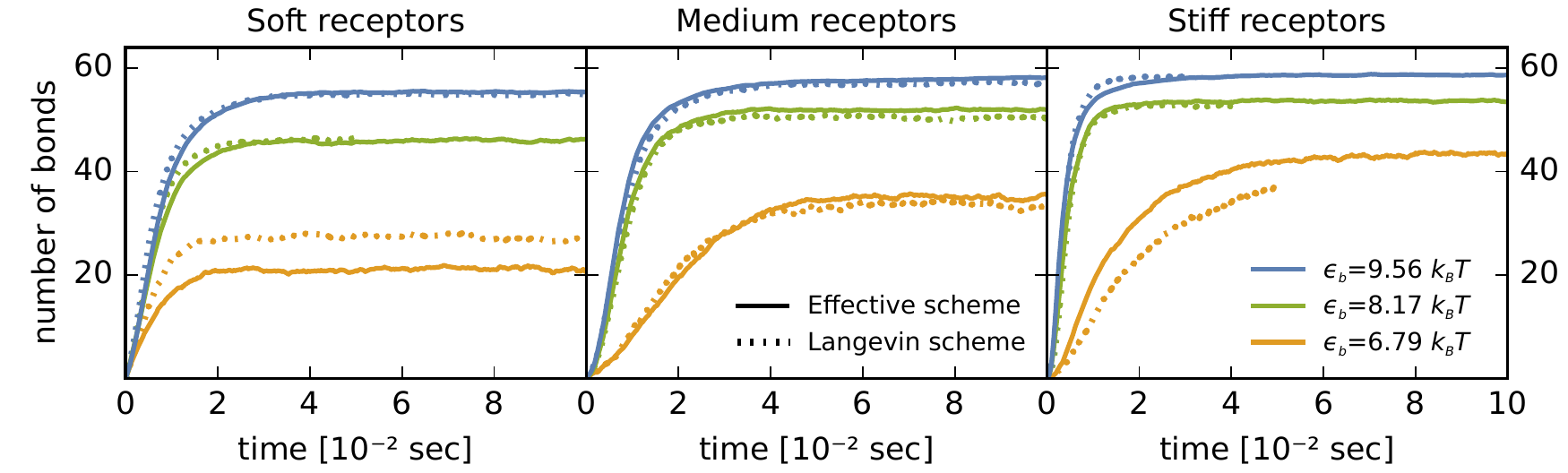}
\caption{Time evolution of the number of bonds for  $\lambda=7.5\times 10^{-3}\, {\kbt}/{\text{nm}^{2}}$ and $k_0=1.6 \times 10^5\, \text{s}^{-1}$(left), $\lambda=2\times 10^{-2}\, {\kbt}/{\text{nm}^{2}}$ and $k_0= 10^5\, \text{s}^{-1}$ (middle) and $\lambda=5\times 10^{-2}\, {\kbt}/{\text{nm}^{2}}$ and $k_0=6.1 \times 10^4\, \text{s}^{-1}$ (right) (remaining parameters see table \ref{tab:para2}). We compare the effective scheme (full lines) with the Langevin scheme (dotted lines).
\label{fig:compare2}}
\end{figure}

We first focus on the simulation of rare events such as is the nucleation of adhesion domains.  The  number of bonds in such a domain can be calculated explicitly within the capillary approximation \cite{Bihr2012}. Once this number is estimated we perform about 2000 simulations with each method to generate the distribution of nucleation times (Fig.~\ref{fig:nucleation}). Specifically, each simulation is set to start from an equilibrated box with zero bonds. When a cluster of bonds of critical size is formed anywhere in the system, the simulation is interrupted, and the time necessary to achieve this domain size is recorded. 
As shown in Fig.~\ref{fig:nucleation}, excellent correspondence of the coarse-grained and the higher-level simulation approach is obtained for the entire distribution of nucleation times.
This agreement could have been anticipated from the successful comparison of Langevin simulations with the analytic model for the nucleation dynamics of a single seed, based on a simplified version of the here used effective rates \cite{Bihr2012}.
The current, more accurate approach fully validates the concept of the effective rates and enables studies of the early stages of the adhesion process in the regimes that are either not accessible to analytic modeling or are extremely demanding from the computational point of view.
Examples of such regimes, which can be now addressed with ease, are fast nucleation, competitive growth of multiple seeds, or diffusion limited nucleation.

\subsection{Full dynamics}
Encouraged by our success reproducing the nucleation dynamics, we validate the Monte Carlo scheme by reproducing the results of the higher level scheme for the full dynamic adhesion process, i.e. nucleation, growth and saturation to  equilibrium. More specifically, for each set of parameters we perform 200 runs over which we average the dynamic process. This level of accuracy was found previously to produce converged results for the Langevin scheme in thermal equilibrium \cite{Reister-Gottfried2008,Reister2011}.

We first explore the correspondence of the two schemes when the diffusion of ligands is fast ($D=5\times10^7\, \text{nm}^2/ \text{s}$), for soft, moderately stiff, and stiff receptors (Fig. \ref{fig:compare2}). In each graph, the number of bonds as a function of time is presented for three different binding affinities (the smallest being at the phase transition to the unstable adhesion dominated by unbinding, two intermediate affinities, and one large affinity where the unbinding is negligible). 
We find that except for critical fluctuations at the phase boundary ($\Eb=6.79\, k_bT$) \cite{Reister2011}, the two approaches show extremely similar dynamics.


\begin{figure}
\centering
\includegraphics[scale=1.0]{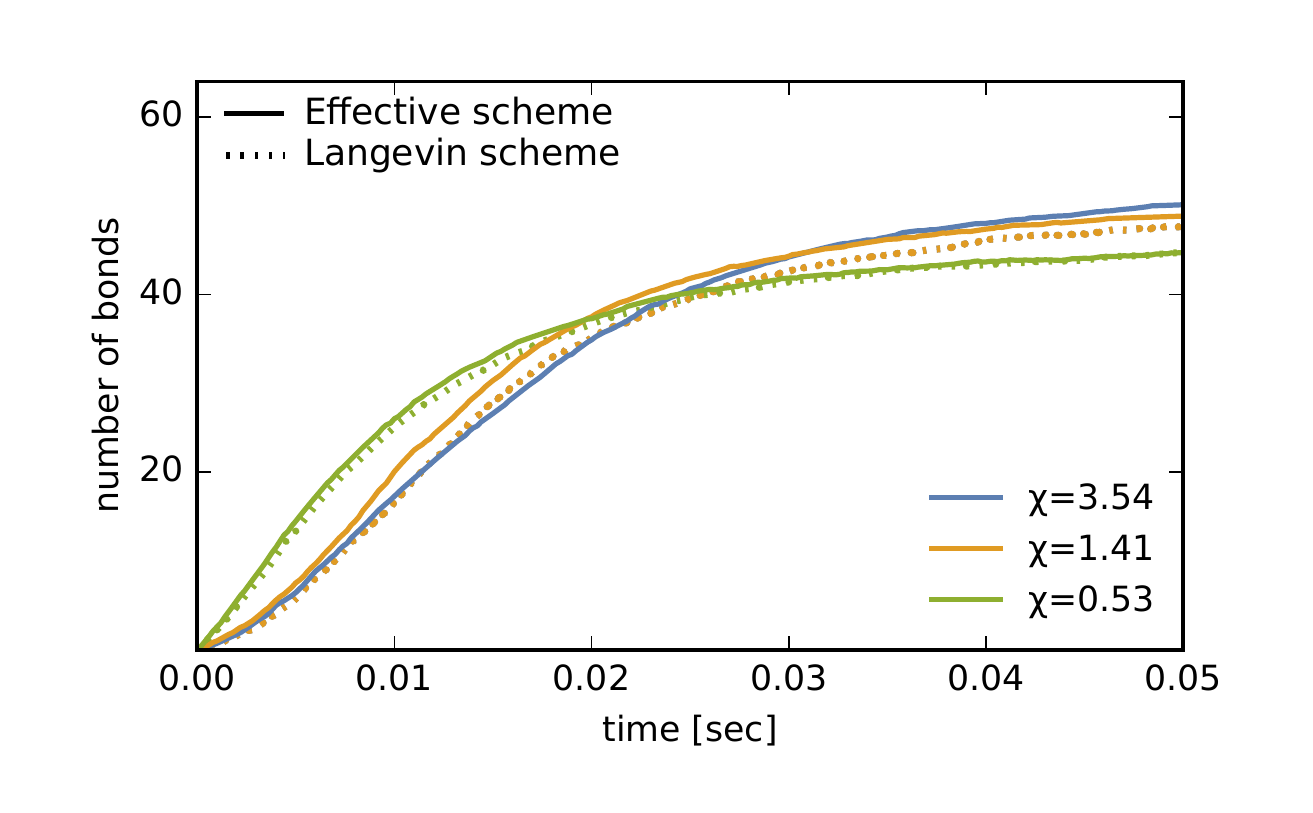}
\caption{
Simulation curves (full lines for effective scheme, dotted lines for Langevin scheme) for different values of  $\chi=\lambda/(8\sqrt{\ka \ga})\quad (\Eb=7.26\kbt$ and  $D=5\times 10^5\, \text{nm}^2/ \text{s}$), remaining parameters as in figure \ref{fig:compare2}. 
\label{fig:chi}
}
\end{figure}

Very similar results are obtained for slow diffusion of ligands (Fig. \ref{fig:chi}), where the adhesion dynamics is shown for three receptor rigidities,  at an intermediate binding affinity.  Equally good, quantitative agreement  is obtained for all affinities  above the transition energy (data not shown). These exceptional results validate the concept of effective rates, and establishes the Monte Carlo approach as a reliable and versatile method for the simulation of protein mediated membrane interactions. It should be noted that different effective Monte Carlo schemes, based on the integration of membrane fluctuations in the Hamiltonian were successful in comparison with the Langevin simulation \cite{Speck2010}, with the time scale of reactions being a free fitting parameter. However, the accuracy of that scheme relied  on the magnitude of the effective cooperativity parameter $\chi$ to be much smaller than one. This dimensionless parameter evaluates the fluctuations of the unbound membrane with respect to the fluctuations of free receptors 
\begin{equation}
\chi\equiv \frac{\lambda}{8 \sqrt{\ka \ga}}.
\end{equation}
The accuracy of the current scheme  does not depend on the effective cooperativity parameter, which for the systems shown in Fig 9 range from 0.53, for the softest receptors, to 3.54, for the stiff receptors.
Actually, the regime of large effective cooperativity parameters seem to be very important in the context of experiments with cells or vesicles  \cite{Smith2010a}.

\section{Simulations of radially growing domains}

One of the basic mechanisms for the growth of adhesion domains is their radial expansion from a stable nucleus.
As observed both in the cellular and cell-mimetic context, with different ligand-receptor pairs, such growth occurs naturally in membranes where the characteristic nucleation time is small compared to the dynamics  of the domain expansion, and is common in situations where one of the binding partners is immobilized \cite{Cavalcanti-Adam2007, Smith2008}.  Particularly well-studied are radially growing domains in ligand-decorated vesicles  binding on a substrate functionalized with receptors \cite{Boulbitch2001,Bihr2014b}.
In these systems, radial growth was used for the determination of the effective binding rate of various ligand-receptor pairs. This rate was found to depend significantly on the properties of the membrane due to strong correlations between the bonds \cite{Bihr2014b}. 

The analysis of the growth dynamics \cite{Boulbitch2001,Freund2004,Gao2005} reveals that the growth of the domain is diffusion limited and the area of the domain increases linearly in time if the concentrations of ligands is smaller than the concentration of receptors  \cite{Shenoy2005, Bihr2014b}.
Otherwise, the growth is reaction limited, and the area grows quadratically. Treating the growth dynamics as a diffusion-reaction problem, the diffusion constant of ligands, and the effective binding constant was extracted from the data \cite{Boulbitch2001}. However, very little is known about the relation of such macroscopic measurements with the underlying microscopic binding and unbinding events, as well as protein motions in the membrane.

Unfortunately, the limited size of systems that can be studied with the Langevin scheme makes this approach unsuited for the analysis of the radial growth process. Nevertheless, using a large number of replicas to reconstruct the representative dynamics, effective affinity, as well as the growth patterns could be identified in the reaction limited case \cite{Reister-Gottfried2008}. However, the issue of the system size is particularly acute for the diffusion limited processes, when a depletion zone around the growing domain forms, and extends  faster than the domain itself \cite{Boulbitch2001,Freund2004,Shenoy2005, Bihr2014b}. This regime, as well as the continuous dynamics in the reaction limited case can only be obtained with the effective Monte Carlo approach developed here. As it will be shown in this section, such a study should clarify how the cooperative effects transmitted by the membrane affect the microscopic rates and the overall dynamics. 
 
\subsection{Simulation details}

We perform a series of Monte Carlo simulations, where we use two opposing square grids of a size of 40.96$\times$40.96$\,$\SI{}{\micro\meter^2} in the diffusion limited case and of 10.24$\times$10.24$\,$\SI{}{\micro\meter^2} in the reaction limited case (typical sizes of a giant unilamellar vesicle). The first grid carries $2.5\times10^5$ receptors (soft or stiff), immobilized on a lattice. To simulate diffusion or reaction limited growth, the second grid is decorated by randomly placing  $5\times10^4$  diffusing ligands or placing immobile ligands above the receptors, respectively. These concentrations, as well as  the other parameters parameters are strongly inspired by the analogous experimental realizations of the system \cite{Bihr2014b}. Specifically,  the height of the membrane ($h_0-l_0=55\,\text{nm}$), curvature of the nonspecific potential ($\gamma=3.125\times 10^{-3}\,{\kbt}/{a^4}$), bending rigidity of the membrane ($\kappa=10\,\kbt$), binding affinity ($\epsilon_b=10\,\kbt$), intrinsic reaction attempt frequency ($k_0=10^5\,\text{s}^{-1}$) and the diffusion constant ($D=$\SI{5}{\micro\meter^2/\second}) is chosen such that the nucleation of domains and the unbinding of bonds are rare. Furthermore, we investigate the reaction and diffusion limited growth regimes for stiff ($\lambda=5\,\kbt/a^{-2}$) and soft receptors ($\lambda=2\,\kbt/a^{-2}$) mimicking bulky cell adhesion receptors and glycoprotein receptors, respectively.

\begin{figure}
	\begin{tabular}{|r|r|}
	\hline
	{\large \textbf{Stiff receptors}}$\quad\quad\quad\quad\quad\,$	&	\large \textbf{Soft receptors}$\quad\quad\quad\quad\,\,\,$\\
	\hline
	\\
	\includegraphics[scale=0.55]{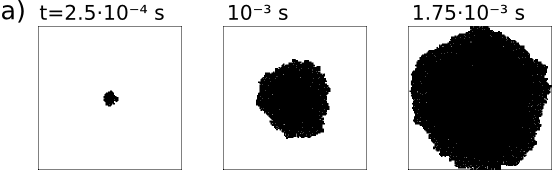}	&
	\includegraphics[scale=0.49]{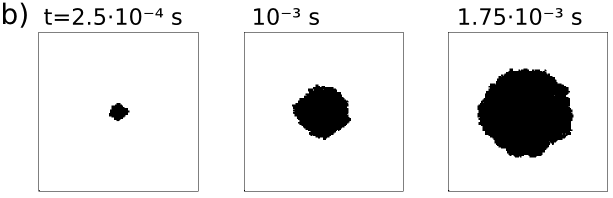}\\
	\includegraphics[scale=1]{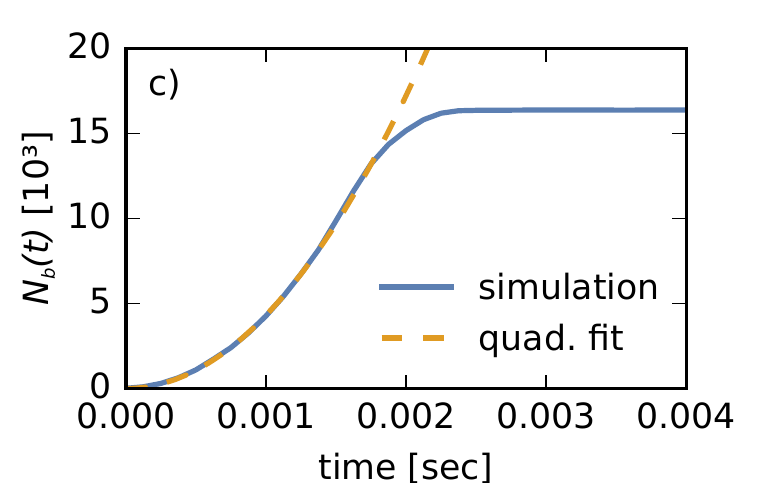}&
	\includegraphics[scale=1]{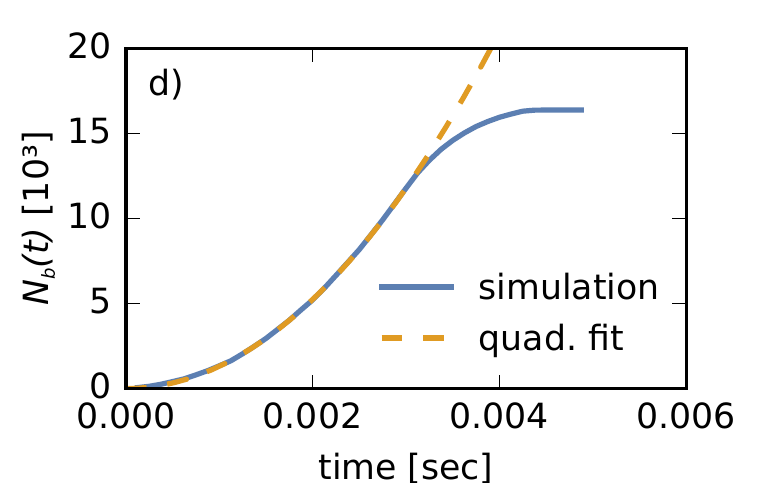}\\
	\includegraphics[scale=1]{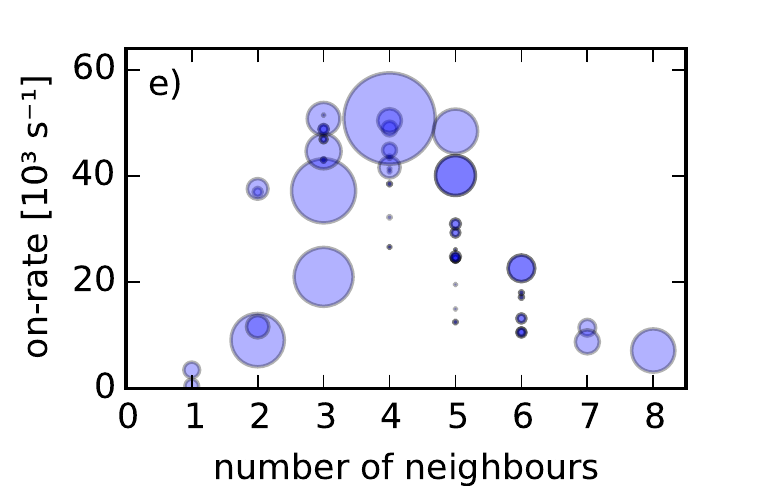}&
	\includegraphics[scale=1]{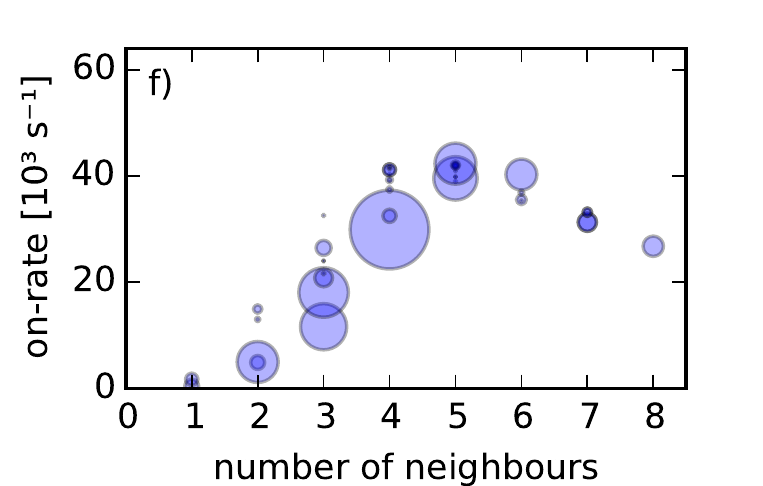}\\
	\includegraphics[scale=1]{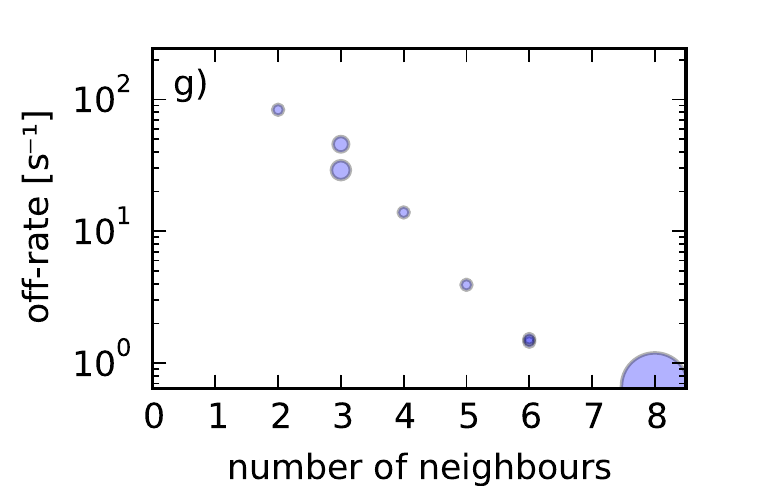}&
	\includegraphics[scale=1]{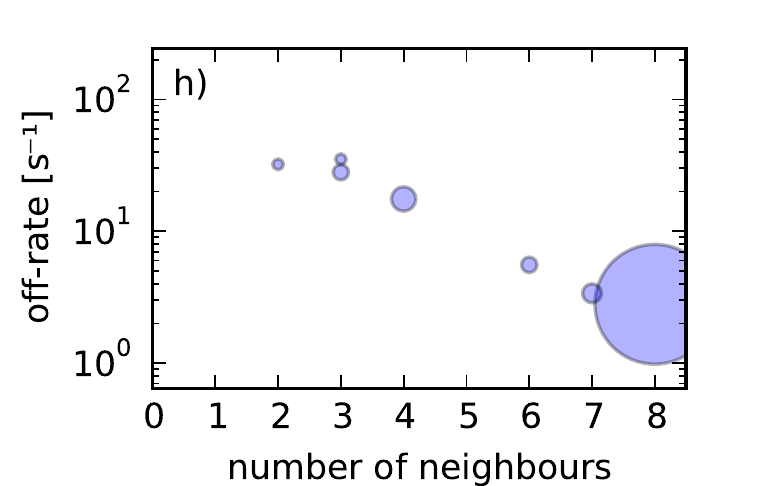}\\
	\hline
	\end{tabular}
	\caption{
	Simulation results of the reaction limited radial growth for stiff receptors (left panels) and soft receptors (right panels).
The first row (a, b) shows snapshots of the growing domain as a function of time whereas the second row (c, d) shows the number of bonds in the domain as a function of time.
In the third (e, f) and the fourth row (g, h), we present bubble charts of the binding and unbinding rates depending on the number of neighboring bonds during the growth phase. 
The area of the bubbles represents the number of reactions with charts.
\label{fig:radial_growth}
	}
\end{figure}

\subsection{Reaction limited growth}

For ligand densities larger than the receptor density, we expect a quadratic growth of the domain area \cite{Boulbitch2001,Cuvelier2003,Bihr2014b} containing $N_b$  uniformly distributed bonds 
\begin{equation}
N_b(t)= \pi ({K}_{R}^{\text{on}})^2 a^4 \rho_0^2 t^2.
\label{eq:most_great}
\end{equation}
Here, $\rho_0$ is the initial density of ligands and ${K}_{R}^{\text{on}}$ is the effective rate at the rim.
This expression explicitly takes into account the two-dimensional nature of the growth process.

The results of our Monte Carlo approach (blue full lines in Fig. \ref{fig:radial_growth} c and d) confirm that the growth of the domain is, quadratic as expected. This is confirmed  by the excellent agreement of the data with fit by equation (\ref{eq:most_great}), shown in Fig. \ref{fig:radial_growth} with dashed orange lines. The observed processes show that growth is faster for stiff (${K}_{R}^{\text{on}}$=\SI{3.7e4}{\second^{-1}}) than for soft (${K}_{R}^{\text{on}}$=\SI{2.0e4}{\second^{-1}}) receptors, presumably because of stronger correlations between bonds. 
Clear deviations from the quadratic behavior take place when the finite size effects start to play a role and the domain begins to cover the whole simulation box.

In order to relate the rates extracted from the fit to the microscopic rates which were actually used to grow the domains, we construct bubble charts for binding and unbinding rates (Fig. \ref{fig:radial_growth} e-h), which are classified by the number of neighbors.
A fixed number of neighbors can be organized in several different configurations around the receptor of interest, which results in the multiple bubbles for each number of neighbors.
In the bubble charts, the area of the bubble is associated with the occurrence of a particular rate in the simulation. 

Interestingly, the effective rate ${K}_{R}^{\text{on}}$ corresponds very well to the average rate recorded in the simulation. Actually, for the stiff bonds the average rate at the rim,
obtained by averaging all rates forming with up to five neighbors $\bar{K}^{\text{on}}$, is $\bar{K}^{\text{on}}$=\SI{3.7e4}{\second^{-1}}, and for
soft bonds $\bar{K}^{\text{on}}$=\SI{2.5e4}{\second^{-1}}. Rates for the formation of bonds with three to five bonds in the neighborhood are most commonly observed (largest bubbles), which is consistent with the formation of new bonds at the edge of the domain.
The rates for the formation of bonds with six or more neighbors are considered to be the results of events from rebinding within the domain, in agreement with the large number of dissociation events with seven and eight neighbors (Fig. \ref{fig:radial_growth} g, h). 

The analysis of microscopic rates in the bubble plots shows that the binding rates have a tendency to increase up to five neighbors.
This happens because the formation of additional bonds, in principle, reduces the distance between the receptor and the ligand at the position of the binding site. The rates for forming the bond with 3-5 neighbors are significantly larger for stiff receptors, which is the source of the difference in the speed of the overall growth process of the domain. The reason for this difference is that for stiff receptors, the membrane approaches closer to the substrate than for soft receptors, which themselves deform while forming a bond, leaving the membrane at a larger height. Rates for forming a bond with 6-8 neighbors decrease with increasing the number of adjacent bonds. This effect is more significant for stiff receptors, because the fluctuations in the membrane are suppressed to a larger extent, and while the distance from the receptor is relatively small, stronger thermal membrane excitation is necessary to bring the ligand into the reaction zone of the receptor.   
The unbinding rates occur less frequently. The most common unbinding rate is the one with 8 adjacent bonds, which is clearly associated with unbinding within the domain.  The unbinding rates decrease exponentially as a function of the number of neighbors, for both stiff and soft receptors, showing the stabilization effects that binding in the surrounding has on the respective bonds.

\subsection{Diffusion limited growth}

\begin{figure}
	\begin{tabular}{|r|r|}
	\hline
	{\large \textbf{Stiff receptors}}$\quad\quad\quad\quad\quad\,$	&	\large \textbf{Soft receptors}$\quad\quad\quad\quad\,\,\,$\\
	\hline
	\includegraphics[scale=1]{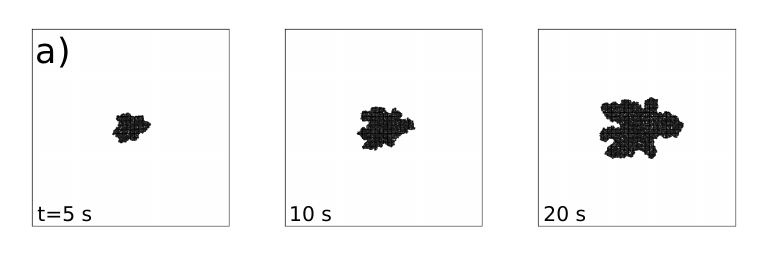}&
	\includegraphics[scale=1]{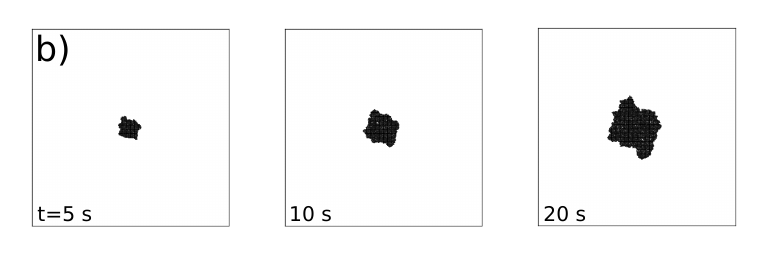}\\
	\includegraphics[scale=1]{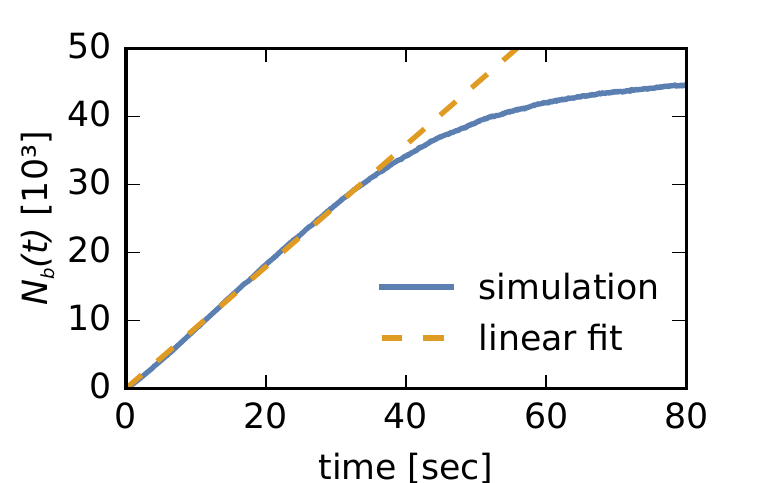}&
	\includegraphics[scale=1]{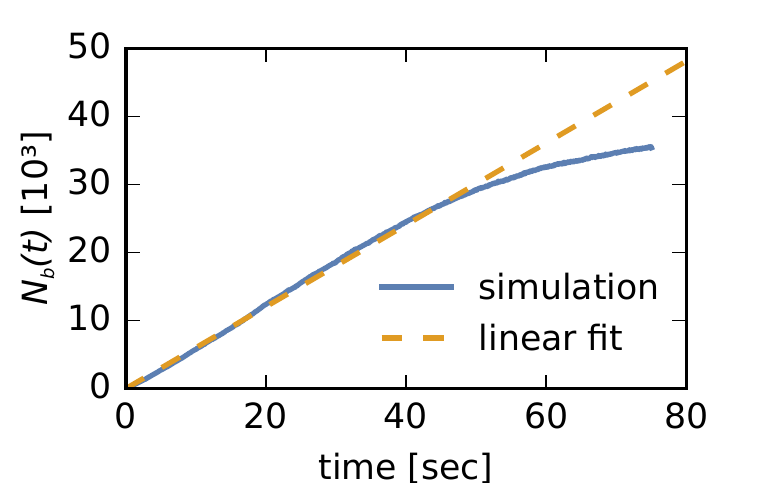}\\
	\hline
	\end{tabular}
	\caption{
	Simulation results of radial growth in the diffusion limited case.
	Top: Snapshots of the growing domains.
	Bottom: Growth curves with linear fit indicating diffusion limited growth.
	Parameters like in Fig. \ref{fig:radial_growth} except for initial ligand density (only $0.4\times10^6$ diffusing ligands).
	\label{fig:diff}}
\end{figure}

For ligand densities lower than the receptor density, the growth is diffusion limited (Fig. \ref{fig:diff}), and depends only implicitly on the effective reaction rates through the density of ligands and bonds at the edge of the domain. In other words, the growth  explicitly depends only on the diffusion constant, and the area of the growing domain $A(t)$ is given by \cite{Boulbitch2001,Shenoy2005}
\begin{equation}
A(t)=4\pi\alpha^2 Dt.
\end{equation}
In this equation, $\alpha$ is a dimensionless speed factor, which is, in two dimensions, determined from the implicit equation \cite{Shenoy2005}  
\begin{equation}
\frac{\rho_0-\rho_e}{\rho_b}=\alpha^2 \exp{\left( \alpha^2 \right)}  \Ei{\left(\alpha^2\right)} .
\label{eq:alpha}
\end{equation}
Here, $\rho_e$ is the density of ligands at the edge of the domain and $\Ei{\left(x\right)}$ is the so-called exponential integral \cite{Gradshteyn2007}.
This relation is obtained from the binder conservation at the rim of the domain and the respective solution of the diffusion equation (see Shenoy and Freund \cite{Shenoy2005} for details).

We numerically solve equation (\ref{eq:alpha}) using the densities of bonds and ligands at the edge of the domain evaluated from the radially averaged density profiles. We obtain $\alpha=0.34$ for stiff receptors and $\alpha=0.27$ for soft receptors. The difference in the speed factors emerges from the somewhat larger density of bonds at the rim of a domain with soft receptors.
Using these speed factors, we can calculate the expected diffusion constant from the linear fit (orange dashed lines in Fig. \ref{fig:diff}). Specifically, we obtain a diffusion constant of \SI{4.8}{\micro\meter^2/\second} for the large bond stiffness, and \SI{5.2}{\micro\meter^2/\second} for the low bond stiffness (right column of figure \ref{fig:radial_growth}) which is is consistent with the diffusion constant in the simulation (\SI{5.0}{\micro\meter^2/\second}). 

\subsection{Remarks in the experimental context}

The obtained results from simulations of the growth of ligand-receptor domains show that it is, in principle, possible to relate macroscopic measurements with the underlying microscopic processes.  From diffusion limited processes we can extract the diffusion constant with excellent accuracy, which is also accessible from experimental data. However, as noted before \cite{Bihr2014b}, issues may arise if the crossover to the saturation of the growth curve due to the finite size of the vesicle or cell occurs relatively quickly and the vesicle runs out of free binders. Furthermore, it is possible to relate the mean reaction rate to the microscopic events.

\section{Conclusions}

We presented two different approaches for simulating protein-mediated adhesion between membranes. 
The first simulation scheme considers the deformation and the fluctuations of the membrane explicitly, by evolving the membrane profile with the help of a Langevin equation. 
The latter was derived from the Helfrich Hamiltonian and included the  hydrodynamic interaction between membrane and  surrounding fluid. The binding and unbinding of ligands and receptors is modeled by Dembo's rates that are in detailed balance with the instantaneous shape of the membrane.
Simpler variants of this scheme have been used  successfully in earlier studies to describe thermal equilibrium \cite{Reister2011} and reaction limited dynamics \cite{Reister-Gottfried2008}.
However, this scheme fails to describe the dynamics on longer length scales as well as diffusion limited processes. The problem arises from the fact that time step is as short as $10^{-9}$ s to correctly recover the membrane thermal excitations. Furthermore, the calculation of the membrane profile requires the use of Fast Fourier Transformations which scale the simulation time with $N\log(N)$, where $N$ is the number of considered membrane segments. As a result, only membrane patches of about $\mu \text{m}^2$ carrying about 1000 proteins can be simulated for about 0.1 seconds.

We overcome these constraints by constructing an effective Monte Carlo scheme. In this scheme, we coarse-grain the adhesion dynamics by integrating the effects of the membrane into a set of effective reaction rates for ligand-receptor (un)binding. These rates are derived by averaging Dembo's rates  over the  membrane height fluctuations, which we do semi-analytically for an arbitrary bond configuration. This allows us to circumvent the explicit treatment of the membrane, and use a much larger time step in the simulation. Consequently, cell-sized objects ($10^4\, \mu \text{m}^2$)  carrying $10^6$ proteins can be simulated for several tens of seconds with the resolution of 10 nm and $10^{-6}\, \text{s}$. In this scheme, the simulation time scales linearly with the number of binders and the simulation time is thus reduced by a factor of about 10\textsuperscript{6} for the parameters used in this study compared to the Langevin approach. 
As shown by an in-depth analysis of the correspondence between the Langevin and Monte Carlos simulations, this increased efficiency is achieved basically without loss of accuracy from the nucleation of  adhesion domains and the early stages of growth to the asymptotic growth behavior and the saturation to an appropriate equilibrium state.

This outstanding performance allows a successful study of completely realistic scaffolding processes. As an example, we performed an analysis for radially growing domains, which is one of the most common scenarios for the development of adhesions. We demonstrate that the measurables that can be extracted from the macroscopic development of the domain can be related to underlying microscopic stochastic processes, namely the protein diffusion and the binding kinetics.    

The simulations presented herein set a foundation for an in-depth analysis of protein transport and complexation dynamics in membranes, which is key to the understanding of the formation of functional microdomains and rafts. Furthermore, processes which present slow convergence or require correlations and signaling on the level of the entire cell are  within the reach of accurate modeling. Now that the adhesion on the level of the membrane can be studied in great detail, the challenge becomes to couple the membrane to other cell structures and processes, which is a direction  for future development.

\ack
We thank D. Schmidt for stimulating discussions and critically reading the manuscript, and M. Knoll for his support in the initial stages of this project. A. S. S. and T. B. acknowledge the funding of the  ERC StG 2013-337283 "MembranesAct", and the DFG RTG 1962 "Dynamic Interactions at Biological Membranes - from Single Molecules to Tissue" at FAU Erlangen.

\appendix
\setcounter{section}{0}
\section{Calculation of the membrane height distribution \label{sec:mhd}}
The membrane height distribution depends on the bond configuration of the membrane  as well on the position of the binding site as can be seen in following equation.
By definition 
\begin{equation}
 p(h(\mathbf{r}))=\funkintegral{}{}{[h'(\mathbf{r})]}p[h'(\mathbf{r}) ]\delta ( h'(\mathbf{r})-h(\mathbf{r})),
\end{equation}
where we have on the left side the probability distribution of the height $p(h(\mathbf{r}))$ at the binding site $\mathbf{r}$, whereas on the right side $p[h'(\mathbf{r}) ]$ is the probability for having a membrane profile $h'(\mathbf{r})$. This probability depends on the bond configuration (i.e. the positions $\mathbf{r_i}$ of the ligand-receptor bonds).
For simplicity, we set $\beta \equiv (\kbt)^{-1} \equiv 1$.
To evaluate the above integral, the Boltzmann weight for $p[h'(\mathbf{r})]$ determined by the Helfrich-Hamiltonian (\ref{eq:helfrich}) is plugged in and the Dirac function is written in the Fourier representation, which gives
\begin{equation}
\begin{split}
p(h(\mathbf{r}))&\propto \integral{}{}{\nu}\funkintegral{}{}{[\hq]}
                  \exp \biggr[
			  -\Summe{i=1}{\NB}\frac{\lambda}{2}\left( h'(\mathbf{r_i}) - \lnull \right)^2\\&-\frac{1}{2A}\Summe{\mathbf{q}}{}\|\hq\|^2 \left(\ka q^4+\ga\right)
                         +i \nu \left(h'(\mathbf{r})-h(\mathbf{r})\right)
                        \biggr].
\end{split}
\end{equation}
We now apply successively $\NB$ Hubbard-Stratonovich transformations, one for each bond term in the sum over $i$. This produces $\NB$ Gaussian integrals over auxiliary fields $\phi_i$. Furthermore, we write $h(\mathbf{r})$ in the Fourier representation. As a result, we get
\begin{equation}
\begin{split}
p(h(\mathbf{r}))\propto&\integral{}{}{\nu}\funkintegral{}{}{[\hq]}
		  \left(\prod_j \integral{}{}{\phi_j}\right) \times\\&
		   \times \exp \biggr\{
		      -\Summe{j}{}\frac{\phi_j^2}{2\lambda} + \Summe{j}{}i \lnull\phi_j
		      -i\Summe{j}{}\frac{\phi_j}{A}\Summe{\mathbf{q}}{} \hq \exp{\left(i \mathbf{q} \cdot \mathbf{r_j}\right)}\\&
		      -\frac{1}{2A}\Summe{\mathbf{q}}{}\|\hq\|^2 \left(\ka q^4+\ga\right)
              + i \nu \frac{1}{A}\Summe{\mathbf{q}}{}\hq \exp{\left(i \mathbf{q} \cdot \mathbf{r}\right)} - i \nu h(\mathbf{r})
		   \biggr\}.
\end{split}
\end{equation}
Integration over $\hq$  is Gaussian integral and leads to 
\begin{equation}
 \begin{split}
    p(h(\mathbf{r}))\propto& \integral{}{}{\nu}
		 \left( \prod_j \integral{}{}{\phi_j}\right)
		  \exp \biggr(
		      -\Summe{j}{}\frac{\phi_j^2}{2\lambda} - i \nu h(\mathbf{r}) + \Summe{j}{}i \lnull\phi_j
		  \biggr)
		  \\& \exp \Biggr\{
		      -\frac{1}{2A}\Summe{\mathbf{q}}{}
		      \Biggr[ \left(
			     \Summe{j}{}        \exp{\left(i \mathbf{q}\cdot \mathbf{r_j}\right)} \phi_j   -\nu  \exp{\left( i \mathbf{q} \cdot \mathbf{r}\right)}
			     \right)\\&
			     \frac{1}{\ka q^4 + \ga}
			    \left(
			     \Summe{k}{}        \exp{\left(-i \mathbf{q}\cdot \mathbf{r_k}\right)} \phi_k   -\nu  \exp{\left( -i \mathbf{q} \cdot \mathbf{r}\right)}
			     \right)
                      \Biggr]
		  \Biggr\}.
 \end{split}
 \label{eq:tra}
\end{equation}
In the following step, the terms within the curly brackets of equation (\ref{eq:tra}) are reorganized, and the sums over $\mathbf{q}$ are converted to integrals that can be evaluated independently leading  to Kelvin functions \cite{Speck2010}. After some sorting, we obtain
\begin{equation}
\begin{split}
    p(h(\mathbf{r}))\propto &   \integral{}{}{\nu}
		  \left(\prod_j \integral{}{}{\phi_j}\right)\exp  \biggr[
		      -\frac{1}{2}\Summe{jk}{} \phi_j \underbrace{\left ( \frac{\delta_{jk}}{\lambda} -\frac{\kei{\left(\q|\mathbf{r_j}-\mathbf{r_k}|\right)}}{2 \pi \sqrt{\ka \ga}}\right)}_{\equiv M_{jk} } \phi_k \\ &   + \Summe{j}{}i \lnull\phi_j
		      \biggr]
		     \exp\left[
			  -\frac{1}{2} \frac{\nu^2}{8 \sqrt{\ka \ga}} - \nu \left(  i h(\mathbf{r})  +\Summe{j}{} \phi_j \frac{\kei{\left(\q|\mathbf{r_j}-\mathbf{r}|\right)}}{2\pi \sqrt{\ka \ga} }            \right)
		      \right],
\end{split}
\end{equation}
where
\begin{equation}
\q\equiv \zeta^{-1}\equiv\sqrt[4]{\frac{\ga}{\ka}}
\label{eq:qnull}
\end{equation}
is the inverse of the lateral correlation length. Performing the Gaussian integrals in $\nu$ gives after some algebra
\begin{equation}
 \begin{split}
    p(h(\mathbf{r}))\propto & \left(\prod_j \integral{}{}{\phi_j}\right) \exp{  \biggr[
				    -\frac{1}{2}  \Summe{jk}{} \phi_i \underbrace{\left( M_{jk} - \frac{2 \kei \left( \q |\mathbf{r}-\mathbf{r_j}|\right)\kei \left( \q |\mathbf{r}-\mathbf{r_k}\right)}{\pi^2 \sqrt{\ka \ga}  } \right)}_{\equiv G_{jk}(\mathbf{r})} \phi_j
				\biggr]}
				\\ & \exp{ \left[
				  i \Summe{j}{}\phi_j \left(\lnull +  h(\mathbf{r}) \frac{4 \kei{ \left(\q |\mathbf{r}-\mathbf{r_j}|     \right)}}{\pi} \right) -\frac{1}{2} 8 \sqrt{\ka \ga} h^2(\mathbf{r})
				\right]}.
\end{split}
\label{eq:GIJ}
\end{equation}
Since the remaining integrals are again Gaussian, one finally gets
\begin{equation}
\begin{split}
p(h(\mathbf{r}))\propto &
		\exp  \Biggr[
			-h(\mathbf{r}) \Summe{ij}{}\lnull G_{ij}(\mathbf{r})^{-1}\frac{4 \kei\left( \q \ |\mathbf{r} - \mathbf{r_j}|\right)}{\pi}\\ &
			-\frac{1}{2}\left(8 \sqrt{\ka \ga} +\Summe{ij}{}
			\frac{16 \kei\left( \q \ |\mathbf{r} - \mathbf{r_i}|\right)G_{ij}(\mathbf{r})^{-1}\kei\left( \q \ |\mathbf{r} - \mathbf{r_j}|\right)}{\pi^2}  \right) h^2(\mathbf{r})
		\Biggr].
\end{split}
\label{eq:notcomsqrt}
\end{equation}
As the probability distribution (\ref{eq:notcomsqrt}) is itself a Gaussian distribution, again, the average height can be calculated by completing the square in the exponent. Consequently, one obtains
\begin{equation}
 \label{eq:expect}
 \langle h(\mathbf{r})\rangle\equiv\bar{h}(\mathbf{r})= \frac
			      {-\frac{4}{\pi}  \Summe{ij}{} \lnull G_{ij}(\mathbf{r})^{-1} \kei{\left( \q \ |\mathbf{r} - \mathbf{r_i}|\right)}}
			      {8 \sqrt{\ka \ga}  + \frac{16}{\pi^2}  \Summe{ij}{} \kei{\left( \q \ |\mathbf{r} - \mathbf{r_i}|\right)}  G_{ij}(\mathbf{r})^{-1} \kei{\left( \q \ |\mathbf{r} - \mathbf{r_j}|\right)}    }.
\end{equation}
The fluctuations are simply given by
\begin{equation}
 \label{eq:fluc}
 \begin{split}
   &\langle h^2(\mathbf{r}) \rangle - \langle h(\mathbf{r}) \rangle^2  \equiv \sigma(\mathbf{r})=\\&
   \left(8 \sqrt{\ka \ga} +\Summe{ij}{}
			\frac{16 \kei\left( \q \ |\mathbf{r} - \mathbf{r_i}|\right)G_{ij}(\mathbf{r})^{-1}\kei\left( \q \ |\mathbf{r} - \mathbf{r_j}|\right)}{\pi^2}  \right)^{-1}.
 \end{split}
\end{equation}

\section*{References}
\bibliographystyle{iopart-num}
\bibliography{doklit.bib}

\end{document}